\documentclass[12pt,a4paper,final]{iopart}

\usepackage{iopams}
\usepackage{cite}

\expandafter\let\csname equation*\endcsname\relax
\expandafter\let\csname endequation*\endcsname\relax

\usepackage[T1]{fontenc}
\usepackage{amsmath}
\usepackage{iopams}
\usepackage{graphicx}
\usepackage[breaklinks=true,colorlinks=true,linkcolor=blue,urlcolor=blue,citecolor=blue]{hyperref}

\def\s{\mathbf{s}}
\def\x{\mathbf{x}}
\def\pa{{\partial \Omega}}
\def\ve{\varepsilon}
\def\M{\mathcal{M}}
\def\P{\mathbb{P}}
\def\C{\mathbb{C}}
\def\R{\mathbb{R}}
\def\K{\mathbf{K}}
\def\Mu{\mathbf{M}}
\def\T{\mathcal{T}}
\def\I{\mathbb{I}}
\def\E{\mathbb{E}}
\def\Z{\mathbb{Z}}

\def\S{\mathcal{S}}

\newcommand{\clr}{\color{black}}

\begin{document}

\title[First-reaction times with target {\clr regions} on boundaries of shell-like
{\clr domains}]{Distribution of first-reaction times with target {\clr regions} on
boundaries of shell-like {\clr domains}}

\author{Denis S Grebenkov$^{1}$, Ralf Metzler$^{2}$ \& Gleb Oshanin$^3$}
\address{$^1$Laboratoire de Physique de la Mati\`{e}re Condens\'ee (UMR 7643),
CNRS -- Ecole Polytechnique, IP Paris, 91120 Palaiseau, France}
\address{$^2$Institute of Physics and Astronomy, University of Potsdam, 14476
Potsdam-Golm, Germany}
\address{$^3$Sorbonne Universit\'e, CNRS, Laboratoire de Physique Th\'eorique
de la Mati\`ere Condens\'ee (UMR CNRS 7600), 4 Place Jussieu, 75252 Paris Cedex
5, France}

\begin{abstract}
We study the probability density function (PDF) of the first-reaction times
between a diffusive ligand and a membrane-bound, immobile imperfect target
{\clr region} in a restricted "onion-shell" {\clr geometry} bounded by two
nested membranes of arbitrary
shapes. For such a setting, encountered in diverse molecular signal transduction
pathways or in the narrow escape problem with additional steric constraints, we
derive an exact spectral form of the PDF, as well as present its approximate form
calculated by help of the so-called self-consistent approximation.  For a particular
case when the nested domains are concentric spheres, we get a fully explicit form of
the approximated PDF, assess the accuracy of this approximation, and discuss various
facets of the obtained distributions. Our results can be straightforwardly applied
to describe the PDF of the terminal reaction event in multi-stage signal
transduction processes.
\end{abstract}

\section{Introduction}

\begin{figure}
\centering
\includegraphics[width=50mm]{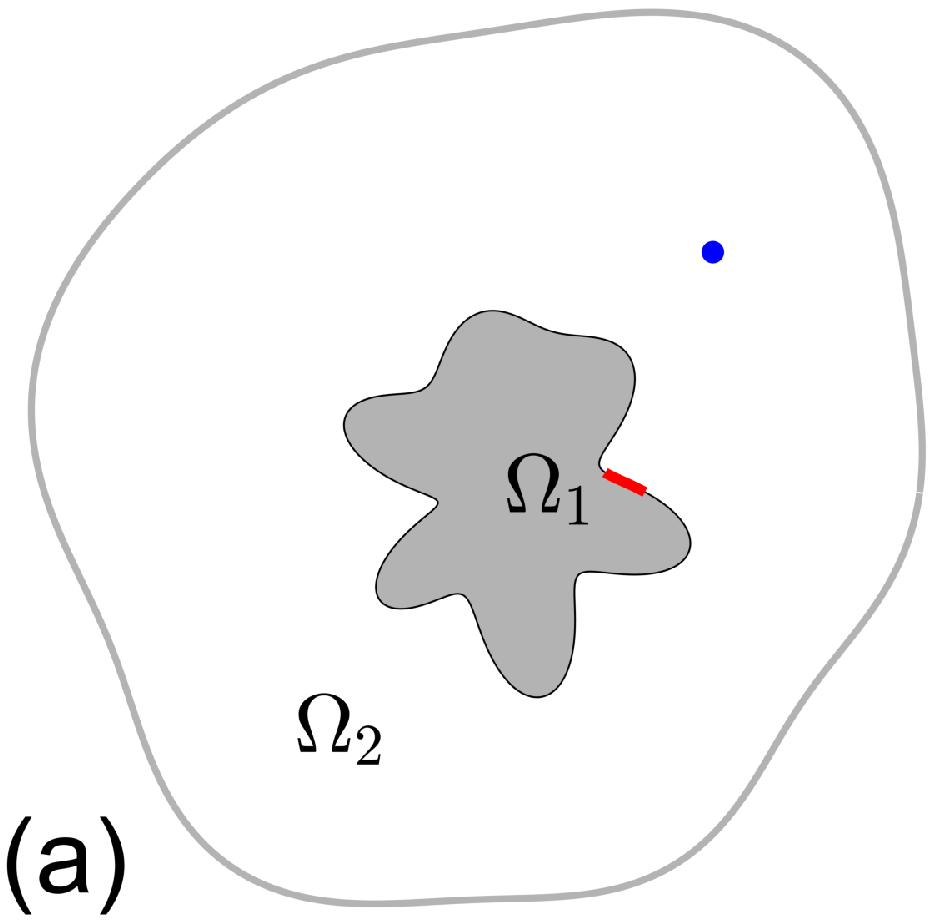} 
\includegraphics[width=50mm]{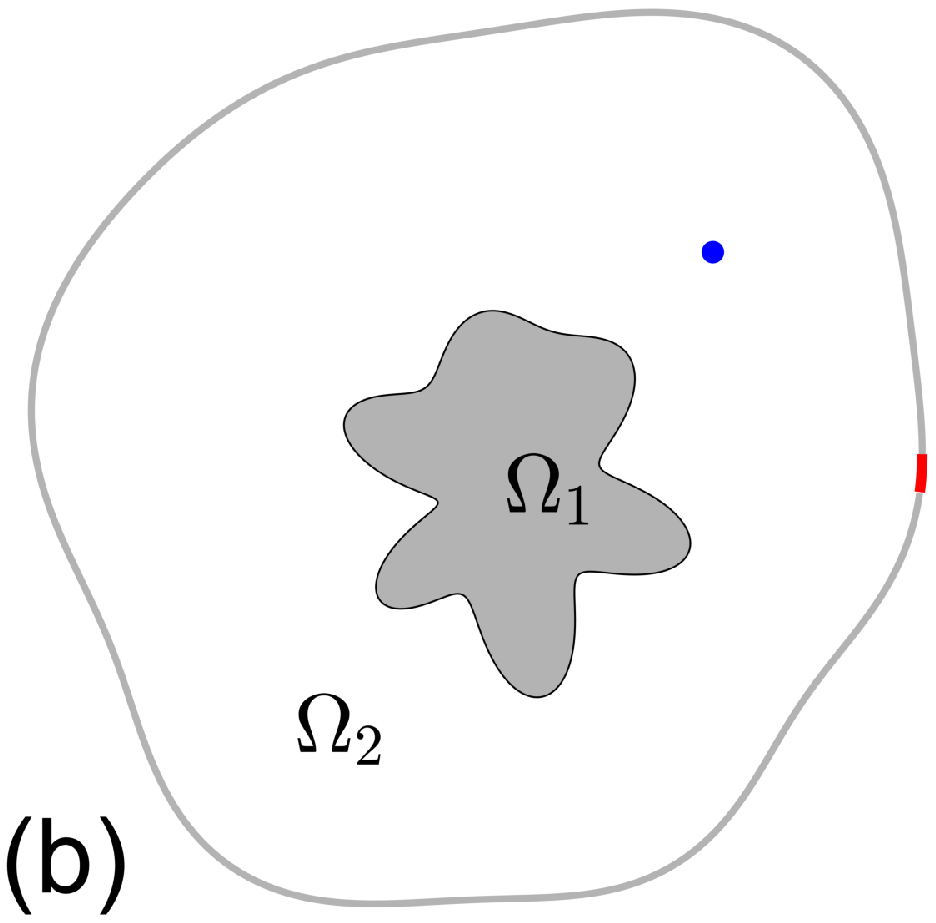}
\caption{Sketch of our geometrical setup. A domain $\Omega_2$ with an
impermeable boundary $\pa_2$ is nesting a smaller domain $\Omega_1$
enclosed by an impermeable boundary $\pa_1$. A particle, whose starting
position is indicated by the blue point, diffuses within the shell-like
{\clr domain} $\Omega=\Omega_2\backslash\Omega_1$ delimited by the two boundaries
and seeks an immobile target (drawn here as an interval in red) placed
on either the inner boundary {\bf (a)} or the outer boundary {\bf (b)}.}
\label{fig:shell}
\end{figure}

A completed reaction event between a diffusive particle and a specific target
site often represents an intermediate yet crucial step in diverse biochemical
and biophysical processes. In many realistic situations a particle diffuses
in a shell-like region delimited by impermeable outer and inner boundaries
and reacts with an immobile target {\clr region (e.g., a catalytic site;
in the remainder, we simply refer to "the target")}
placed on either of the boundaries.
In some applications, this target is located on the inner boundary
(figure \ref{fig:shell}(a)). This is a common situation in chemoreception
processes \cite{berg,lau} as well as, more generally, in cellular signal
transduction pathways \cite{trans1,trans2,alberts}. Here the shell-like {\clr domain}
can be an extracellular medium and the inner boundary represents the (outer)
plasma membrane of a cell. In such a setting the particle is commonly referred to as
the "ligand" or, in the literature on signal transduction, as the first "messenger".
The target is a receptor that undergoes a conformational change when the
first messenger binds to it, stimulating then a synthesis of the second messenger
which moves {\clr inside} the cell itself, i.e., within the inner domain. Similarly the
particle may cross the cell wall through membrane pores. In a different scenario,
the shell-like {\clr domain} can be the intracellular medium (cytoplasm). Then the outer
boundary is the cellular membrane and the particle can be, e.g., the second
messenger, which searches diffusively for a specific target on the inner boundary,
e.g., the nuclear membrane, then launching a cascade of processes upon binding to
this site. In other situations, the target can be located on the outer
boundary (figure \ref{fig:shell}(b)). It can be a tiny aperture---an escape
window, in which case the shell-like {\clr domain delimited by} two boundaries can be regarded
as the cortex region, while the inner domain may represent, e.g., the centrosome,
as studied in \cite{100,100c}. Such a geometrical setup differs from (and is more
complicated than) usually studied geometrical settings of the by-now classical
Narrow Escape Problem (NEP) \cite{5c,7,7b,Caginalp12,Marshall16,Grebenkov16c,
Lindsay17,Bernoff18a,6,6a,8,9,9a,9b,9c,11} due to the presence of a centrosome.
From a different perspective, this case can be viewed as an initial step in
cell-to-cell communication processes \cite{lau,trans1,trans2,alberts} (see also
recent results in \cite{mat,mor}). In this
important situation, a particle is the first signalling molecule emitted at a
specified location within the intracellular medium, which then has to engage
with the target on the cellular membrane. After a reaction with this {\clr region},
the cell secretes a ligand that moves diffusively in the extracellular domain
until it binds to another cell. The binding event is followed by the so-called
internalisation process: the signal propagates within the second cell in a
cascade of successive reactions, as described above.

Understanding the kinetic behaviour of such multi-stage processes comprising
a specific reaction event as one of its controlling factors is impossible
without the knowledge of how long such a single reaction stage lasts,
starting with the launch of a diffusive particle and terminating at the
instant of a successful reaction event. The duration of this
stage is a random variable, which in what follows we call the first reaction
time (FRT). Its distribution can be rather broad even in bounded
systems with a simple geometry \cite{carlos1,carlos2}. From the mathematical
point of view, the derivation of the probability density function (PDF) of FRTs
is well-established \cite{redner,redner1,redner2}: it consists of solving the
diffusion equation within the domain under consideration, subject to the
appropriate boundary conditions. This solution determines the
so-called survival probability, i.e., the probability that the particle did
not react up to the time instant $t$, and the desired FPT PDF is deduced as the
derivative of the survival probability with respect to $t$, taken with the
minus sign. However, the domain may have a complicated shape, and even more
crucially, the appropriate boundary conditions for chemical reactions are
the so-called \textit{mixed} boundary conditions: a zero-current boundary
condition on the impermeable surfaces and a reactive
boundary condition on the target. Thus, in general, this
problem has no {\it explicit exact} solutions, except for asymptotical
results obtained for the NEP in the limit of a
vanishingly small size of the escape window \cite{7,7b,5c} as well as several
spectral exact solutions derived for simple geometries \cite{Grebenkov19a,
Grebenkov19b,Grebenkov20b}. In the general case, one therefore either has to
resort to stochastic simulations {\clr (ranging from lattice random walks 
or basic Monte Carlo schemes to more advanced techniques such as enhanced 
Green's Function Reaction Dynamics (eGFRD) \cite{Sokolowski19}),} or to a
numerical analysis of the boundary value problem via standard discretisation
schemes (such as, e.g., finite difference or finite element methods). {\clr
Here, we follow another direction which consists in developing} approximate
analytical methods. The latter are, of course, more advantageous because they
show how the pertinent properties depend on the system parameters, a dependence
that can be verified by comparison with simulations results {\clr or experimental
data.} The predictions of approximate methods can be also tested against available
exact solutions in simple geometries.

One such approximate method is the self-consistent approximation (SCA) originally
developed in \cite{szabo} to calculate the Smoluchowski-type constant for
reactions with a small centre situated on the otherwise impenetrable surface
of a spherical domain. In essence, {\clr one replaces}
the mixed Robin-Neumann boundary condition by an inhomogeneous Neumann
boundary condition, and then establishes an appropriate closure relation
for the current through the surface of the target. This approach has been
subsequently invoked to study several reaction-diffusion problems. Specifically,
it was used to calculate (i) the mean velocity of a directional motion of a
colloid decorated with a catalytic patch, which prompts reactions in the
embedding medium and creates a self-propulsive force \cite{colloid}; (ii) the
mean FRT in the NEP with long-ranged interactions with the boundary and an
entropic barrier at the aperture \cite{11}; and (iii) the mean FRT for particle
binding to a specific site on a stretched DNA \cite{dist5}. For the latter
case the SCA was also compared against the predictions of another approximate
approach, the so-called boundary homogenisation method, and was shown to be
more accurate than the latter \cite{dist7}. Extending the SCA beyond the mean
rates, the full PDF of the FRT was determined for the NEP in circular and
spherical domains \cite{dist2} and for the binding kinetics to a specific site
on an impermeable cylinder, which mimics an elongated DNA, in a bigger
cylindrical domain \cite{dist4}. Moreover, upon a comparison with the
numerical solution of the mixed boundary-value problem it was demonstrated
that the SCA is a very reliable approximation, whose predictions agree very
well with the numerical results obtained through the finite elements method.

Here we study the statistics of the FRTs in the restricted "onion-shell"
geometry depicted in figure \ref{fig:shell} which consists of two nested
bounded domains: an inner domain $\Omega_1$ placed inside a larger domain
$\Omega_2$. A diffusive point-like particle is launched at time $t=0$ from an \emph{
arbitrary\/} fixed position within the shell-like {\clr domain} $\Omega=\Omega_2
\backslash\Omega_1$ delimited by the impermeable boundaries $\pa_1$ and
$\pa_2$. The particle then searches for the immobile target ${\cal T}$
located at either of the delimiting boundaries. Such settings correspond to
many realistic situations encountered in molecular signal transduction or in
the NEP with additional steric constraints \cite{100,100c}. Considering the
reaction between the particle and the target we pursue the general case
in which a reaction (or binding event) is not perfect and takes place only
with a finite probability. This defines the intrinsic chemical reactivity
$\kappa$ with $0<\kappa<\infty$ (see, e.g., \cite{Collins49,szabo,colloid,
11,redner1,ol1,lawley,lawley2,denis,Grebenkov20}). When $\kappa=\infty$ we
are in the case of perfect reactions, occurring immediately upon first contact.
The FRT in this case is then exactly the first-passage time to the target,
similar to Smoluchowski's original assumption \cite{smoluchowski}. For finite
$\kappa$, the reaction is not instantaneous and may not complete upon the first
encounter of the diffusive particle with the target, thus necessitating
repeated diffusive loops and reaction attempts. Clearly, the FRT is always
longer than the first-passage time and strongly depends on the value of
$\kappa$. Our aim here is to calculate the PDF of the FRT.

We proceed as follows. We start with the general situation in which the nested
domains have arbitrary shape with (sufficiently smooth) impermeable boundaries.
Capitalising on the recent analysis \cite{Grebenkov19a}, we present a formally
exact spectral solution of the problem and {\clr then} develop an SCA for domains of
arbitrary shapes. Note that the {\clr SCA} has only been worked out {\clr previously} for
some particular geometries. Here, we establish a general theoretical framework
which includes previous geometrical settings as particular cases. Moreover, all
steps involved in this general approach are clearly identified and will thus be
useful and instructive for the analysis of the FRT statistics in other systems.
We then apply the developed framework to the case when the domains $\Omega_1$ and
$\Omega_2$ are concentric balls, such that $\Omega$ has the form of a spherical
shell. For this particular case, we present explicit forms of the FRT PDF,
discuss its detailed behaviour and also compare it against the formally exact
spectral solution, in which the entering matrices are inverted numerically {\clr (see below).} We
note that the obtained FRT PDF can be considered as a "building block" in more
complex signal transduction pathways taking place in nested bounded domains
(see, e.g. \cite{race} for more details).

The paper is organised as follows. In section \ref{sec:general}, we present
the mathematical formulation of the FRT problem and its formal spectral
solution as obtained in \cite{Grebenkov19a}. Mainly we derive a general form
of the SCA, which is valid for domains of arbitrary shape and connectivity,
even including unbounded domains with a bounded boundary.\footnote{\clr The
mathematical term "bounded boundary" is used in the sense of a compact
(loosely, "finite") boundary.} In section
\ref{sec:shell} the developed framework is applied to spherical shell domains, for
which we evaluate the novel exact spectral solution and also provide an
explicit prediction for the generating function of the FRT. The result is
also discussed for some particular limiting cases. Finally, we analyse the
corresponding FRT PDF via numerical inversion of the Laplace transform. In
section \ref{sec:discussion} we present a brief discussion of the general
form of the obtained FRT PDF and its asymptotic behaviour. In addition, we
show how the shape of the PDF depends on the local curvature of the boundary
in the vicinity of the target and also on the radius of the inner domain.
We conclude in section \ref{sec:conclusion} with a brief summary of our results
and outline some perspectives for future research. Details of intermediate
calculations and some of the results are relegated to Appendices. In
\ref{sec:comparison} we discuss the accuracy of the SCA in spherical-shell
domains as compared to the exact spectral solution, in \ref{sec:MFPT}
we determine the mean FRT, while in \ref{sec:planar} we sketch an
extension of our results to a planar circular annulus domain.  {\clr
\ref{sec:spread} yields complementary insights via the analysis of the
distribution of reaction event locations.}

\section{Spectral solution and the SCA in Laplace domain}
\label{sec:general}

Consider a point-like particle with diffusivity $D$ which starts at time zero
at position $\x$ and diffuses within a bounded, $d$-dimensional Euclidean
{\clr domain} $\Omega\subset\R^d$. The boundary $\pa$ of $\Omega$ is assumed to be
sufficiently smooth and is reflecting everywhere, except for the target
denoted by $\T$. Upon {\clr hitting} the target, a particle reacts with it
with a finite probability which defines the intrinsic chemical reactivity
$\kappa$ \cite{szabo,colloid,11}. The FRT $\tau$ is a random variable, that
is distributed according to the PDF
\begin{equation}
H(t|\x)=-\frac{\partial S(t|\x)}{\partial t}.
\end{equation}
Here $S(t|\x)=\P_{\x}\{\tau>t\}$ is the survival probability, that satisfies
the diffusion equation \cite{redner}
\begin{equation}
\label{eq:diffusion}
\frac{\partial S(t|\x)}{\partial t}=D\Delta S(t|\x)\quad(\x\in\Omega),
\end{equation}
where $\Delta$ is the Laplace operator, subject to the initial condition 
\begin{equation}
\label{eq:initial}
S(0,\x)=1  
\end{equation}
and the mixed Robin-Neumann boundary conditions
\begin{equation}
\label{eq:boundary}
\left\{\begin{array}{ll}-D\partial S(t,\x)/\partial n=\kappa S(t,\x) & (\x\in\T)\\
-D\partial S(t,\x)/\partial n=0 & (\x\in\pa\backslash\T)\end{array}\right..
\end{equation}
Here $\partial/\partial n$ is the normal derivative oriented outwards from the
domain. The first relation in \eqref{eq:boundary} states that the diffusive flux
toward the target $\T$ is equal to the reaction flux on that target.
In turn, the second relation indicates zero diffusive flux on the
remaining part of the impermeable boundary. These boundary conditions
can be compactly written as
\begin{equation}
\label{eq:boundary2}
-\frac{\partial S(t|\x)}{\partial n}=q\I_{\T}(\x)S(t|\x) \quad (\x\in\pa),
\end{equation}
where $\I_{\T}(\x)$ is the indicator function of the subset $\T$ that is
$\I_{\T}(\x)=1$ if $\x\in\T$ and $0$ otherwise.

{\clr The parameter $q = \kappa/D$ (in units of inverse length) ranges
from $0$ to infinity and quantifies the interplay between bulk
diffusive transport (characterised by $D$) and surface reaction
(characterised by $\kappa$).  After multiplication by an appropriate
length scale $R$ of the confining domain, one can distinguish
reaction-limited (small $qR$) and diffusion-limited (large $qR$)
processes \cite{Sapoval94,Sapoval02,Grebenkov03}.  The inverse of $q$
can also be interpreted as the size of a typical region around the
first arrival point on the boundary, in which the reaction occurs (see
\cite{Sapoval05,Grebenkov06c,Grebenkov15c} and \ref{sec:spread} for
details). }

\subsection{Spectral solution in Laplace domain}

We focus on the Laplace-transform
\begin{equation}
\tilde{H}(p|\x)=\int\limits_0^\infty dt e^{-pt}H(t|\x)
\end{equation}
of the first-reaction time PDF, which is related to the Laplace-transformed survival
probability via $\tilde{H}(p|\x)=1-p\tilde{S}(p|\x)$ and thus satisfies the
modified Helmholtz equation
\begin{equation}
\label{eq:Hp_Helmholtz}
(p-D\Delta)\tilde{H}(p|\x)=0 \quad (\x\in\Omega)
\end{equation}
subject to the mixed boundary conditions
\begin{equation}
\label{eq:boundary3}
\frac{\partial \tilde{H}(p|\x)}{\partial n}+q\I_{\T}(\x)\tilde{H}(p|\x)=q
\I_{\T}(\x) \quad (\x\in\pa). 
\end{equation}
As discussed in \cite{Grebenkov19a} a formal spectral solution of this problem
can be obtained by use of the Dirichlet-to-Neumann operator $\M_p$, which
associates to a given function $f(\x)$ on the boundary $\pa$ another function
$g(\x)$ on that boundary as follows:
\begin{equation}
\label{eq:DtN}
\M_p ~:~ f(\x)\to g(\x)=\left.\left(\frac{\partial u(\x)}{\partial n}\right)
\right|_{\pa}, \mbox{ where }\left\{\begin{array}{ll}(p-D\Delta)u(\x)=0 &
(\x\in\Omega)\\u(\x)=f(\x) & (\x\in\pa)\end{array}\right..
\end{equation}
In other words, for a given function $f(\x)$, one solves the modified Helmholtz
equation with Dirichlet boundary condition $u(\x)=f(\x)$ and then evaluates the
normal derivative of $u(\x)$. It is known that $\M_p$ is a pseudo-differential
self-adjoint operator which represents the action of the normal derivative onto
a solution of the modified Helmholtz equation \cite{Arendt14,Daners14,Arendt15,
Hassell17,Girouard17}. As a consequence the mixed boundary condition
(\ref{eq:boundary3}) can be written as
\begin{equation}
\M_p\tilde{H}(p|\x)+q\I_{\T}(\x)\tilde{H}(p|\x)=q\I_{\T}(\x).
\end{equation}
Keeping the same notation $\I_{\T}$ for the operator of multiplication by the
function $\I_{\T}(\x)$, one can formally invert this operator equation to get
\begin{equation}
\tilde{H}(p|\x)=\bigl(\M_p/q+\I_{\T}\bigr)^{-1}\I_{\T}(\x) \qquad (\x\in\pa),
\end{equation}
from which the solution $\tilde{H}(p|\x)$ can be extended to the whole domain
$\Omega$ by using the Dirichlet Green's function (see below). Since the boundary
$\pa$ was assumed to be bounded, the spectrum of the self-adjoint operator
$\M_p$ with $p\geq 0$ is discrete, i.e., there is an infinite sequence of
eigenvalues, $\mu_0^{(p)}\leq\mu_1^{(p)}\leq\ldots\nearrow +\infty$, associated
to eigenfunctions $\{v_n^{(p)}(\x)\}$ forming an orthonormal basis of the space
$L_2(\pa)$ of square integrable functions on $\pa$. Using these eigenfunctions,
one can represent the above solution as
\begin{equation}
\label{eq:Hp_spectral}
\tilde{H}(p|\x)=\sum\limits_{n,n'=0}^\infty v_n^{(p)}(\x)\biggl[\bigl(\Mu^{(p)}/q
+\K^{(p)}\bigr)^{-1}\biggr]_{n,n'}C_{n'}^*\qquad (\x\in\pa),
\end{equation}
where $\Mu^{(p)}$ is the diagonal matrix of eigenvalues $\mu_n^{(p)}$, $\Mu^{(p)}_
{n,n'}=\delta_{n,n'}\mu_n^{(p)}$ ($\delta_{n,n'}$ denoting the Kronecker symbol),
\begin{equation}
\label{eq:Cn}
C_n=\int\limits_{\pa}d\x\I_{\T}(\x)v_n^{(p)}(\x)=\int\limits_{\T}d\x v_n^{(p)}(\x),
\end{equation}
and
\begin{equation}
\bigl[\K^{(p)}\bigr]_{n,n'}=\int\limits_{\pa}d\x v_n^{(p)}(\x)\I_{\T}(\x)[v_{n'}^{
(p)}(\x)]^*=\int\limits_{\T}d\x v_n^{(p)}(\x)[v_{n'}^{(p)}(\x)]^*
\end{equation}
is the matrix representation of the operator $\I_{\T}$ with respect to
the eigenbasis $\{v_n^{(p)}\}$. In the above the asterisk denotes the
complex conjugate. This construction is an exact
solution of the problem defined by \eqref{eq:Hp_Helmholtz} and
\eqref{eq:boundary3}, {\clr which requires, however, the inversion of the infinite-dimensional matrix $\Mu^{(p)}/q
+\K^{(p)}$.} We stress that it is valid for any Euclidean
domain $\Omega$ with a sufficiently smooth bounded boundary and target
$\T$ of any shape, not necessarily small (it may, in fact, cover
the entire boundary $\pa$), nor necessarily simply-connected; in fact, the
solution holds for multiple target, as well.  

While the solution \eqref{eq:Hp_spectral} was derived for
$\x\in \pa$, its extension to any $\x\in \Omega$ can be obtained in a
standard way by solving the Dirichlet boundary value problem for the
modified Helmholtz equation \eqref{eq:Hp_Helmholtz},
\begin{equation}
\label{eq:auxil12}
\tilde{H}(p|\x)=\int\limits_{\pa}d\s\tilde{H}(p|\s)\underbrace{\left(-D
\frac{\partial}{\partial n}\tilde{G}_\infty(\x',p|\x)\right)_{\x'=\s}}_{
=\tilde{j}_\infty(\s,p|\x)},
\end{equation}
where $\tilde{G}_\infty(\x',p|\x)$ is the Dirichlet Green's function
for equation \eqref{eq:Hp_Helmholtz}. Here, the expression in
parentheses is the Laplace transform of the probability flux density
$j_\infty(\s,t|\x)$, that yields a probabilistic interpretation to
this extension: a diffusing particle first hits the boundary $\pa$ at
some point $\s$ (the first-passage problem described by
$j_\infty(\s,t|\x)$), from which it continues searching the target
(the first-reaction time problem described by $H(t|\s)$). Their
convolution in time domain takes the form of a product in
\eqref{eq:auxil12} in Laplace domain.  Substituting the spectral
decomposition \eqref{eq:Hp_spectral}, one finally gets,
\begin{equation}
\label{eq:Hp_spectral1}
\tilde{H}(p|\x)=\sum\limits_{n,n'=0}^\infty V_n^{(p)}(\x)\biggl[\bigl(\Mu^{(p)}/q
+\K^{(p)}\bigr)^{-1}\biggr]_{n,n'}C_{n'}^*\qquad (\x\in\Omega),
\end{equation}
with
\begin{equation}
\label{eq:Vnp}
V_n^{(p)}(\x)=\int\limits_{\pa}d\s v_n^{(p)}(\s) \tilde{j}_\infty(\s,p|\x). 
\end{equation}
In this way, one extends the eigenfunctions $v_n^{(p)}(\s)$ of the
Dirichlet-to-Neumann operator, defined on the boundary $\pa$, into the
whole domain $\Omega$.  Such extensions can also be understood as the
eigenfunctions of the associated Steklov problem.

At a first glance, the spectral representation \eqref{eq:Hp_spectral} may look
useless, as it expresses an unknown but intuitively appealing quantity $\tilde{
H}(p|\x)$ in terms of several unknown and less clear quantities (eigenfunctions
$v_n^{(p)}$, matrices $\Mu^{(p)}$ and $\K^{(p)}$). In section \ref{sec:shell}
we will discuss that in some geometric settings the latter quantities can be
found explicitly thus rendering the above formal solution suitable for both
numerical computations and analytical studies.

\subsection{Self-consistent approximation}

We now turn to the SCA as developed and applied in \cite{szabo,colloid,11,
dist5,dist7,dist4,dist2}. This approximate method consists in replacing
the mixed Robin-Neumann boundary condition \eqref{eq:boundary2} by an
effective Neumann boundary condition. The latter preserves a zero flux
boundary condition at the reflecting part $\pa\backslash\T$ of the boundary
and stipulates that the current through the target $\T$ is a constant that
does not depend on the spatial coordinates within the target.
In other words, one aims at solving the modified problem
\begin{align}
\label{eq:Sinhom}
(p-D\Delta)\tilde{S}_{\rm app}(p|\x)&=1\quad(\x\in\Omega),\\  
\label{eq:BC_Sinhom}
-\frac{\partial\tilde{S}_{\rm app}(p|\x)}{\partial n}&=\frac{J}{p}\I_\T(\x)
\quad(\x\in\pa),
\end{align}
where the subscript "app" highlights that the solution of this problem,
$\tilde{S}_{\rm app}(p|\x)$, is meant to approximate $\tilde{S}(p|\x)$.
The adjustable parameter $J$ has to be determined from the self-consistent
closure condition, which requires that the original Robin boundary
condition on the target $\T$ is satisfied {\it on average}:
\begin{equation}
\int\limits_{\T}d\x\left(-\frac{\partial\tilde{S}_{\rm app}(p|\x)}{\partial n}
\right)=\int\limits_{\T}d\x\bigl(q\tilde{S}_{\rm app}(p|\x)\bigr).
\end{equation}
Using the modified boundary condition \eqref{eq:BC_Sinhom}, one gets
\begin{equation}  \label{eq:S_closureJ}
J=\frac{qp}{|\T|}\int\limits_{\T}d\x\tilde{S}_{\rm app}(p|\x),
\end{equation}
where $|\T|$ is the area of the target $\T$. In turn, {\clr setting
\begin{equation*}
\tilde{H}_{\rm app}(p|\x)=1-p\tilde{S}_{\rm app}(p|\x) = J \tilde{h}_{\rm app}(p|\x),
\end{equation*}
} the problem defined by equations
\eqref{eq:Sinhom} to \eqref{eq:BC_Sinhom} reads
{\clr
\begin{align} 
\label{eq:Hp_modif}
(p-D\Delta)\tilde{h}_{\rm app}(p|\x)&=0\quad(\x\in\Omega),\\
\label{eq:BC_Hinhom}
\frac{\partial\tilde{h}_{\rm app}(p|\x)}{\partial n}&= \I_{\T}(\x)\quad(\x\in\pa),
\end{align}
while the closure relation \eqref{eq:S_closureJ} becomes
\begin{equation*} 
J=q\left(1-\frac{J}{|\T|}\int\limits_{\T}d\x\tilde{h}_{\rm app}(p|\x)\right),
\end{equation*}
from which
\begin{equation} 
\label{eq:J_Hp}
J=\left(\frac{1}{q} + \frac{1}{|\T|}\int\limits_{\T}d\x\tilde{h}_{\rm app}(p|\x)\right)^{-1} .
\end{equation}
}
The modified boundary value problem, equations \eqref{eq:Hp_modif} to
\eqref{eq:BC_Hinhom}, is generally much simpler than the original one. In
particular, it can be solved in an explicit exact form for some geometric
settings, see the examples in \cite{szabo,colloid,11,dist5,dist7,dist4,dist2} {\clr and in section \ref{sec:shell}.}
In general, one can express its solution by using the eigenbasis of the
Dirichlet-to-Neumann operator. In fact, as $\M_p$ represents the normal
derivative in \eqref{eq:BC_Hinhom}, its inversion immediately yields
\begin{equation}
\label{eq:Happ}
\tilde{H}_{\rm app}(p|\x)=J\M_p^{-1} \I_{\T}(\x)=\sum\limits_{n=0}^\infty
\frac{v_n^{(p)}(\x)}{\mu_n^{(p)}/J}C_n^*\qquad(\x\in\pa),
\end{equation}
where the $C_n$ are given by \eqref{eq:Cn}. Substituting this solution into
\eqref{eq:J_Hp}, we get 
\begin{equation}
\label{eq:J}
J=\left(\frac{1}{q}+\frac{1}{|\T|}\sum\limits_{n=0}^\infty\frac{|C_n|^2}{\mu_n^{
(p)}}\right)^{-1}.
\end{equation}
We emphasise that the solution \eqref{eq:Happ} with \eqref{eq:J} does not require
any matrix inversion and is thus explicit, once the eigenbasis of the
Dirichlet-to-Neumann operator is known {\clr (e.g., in the cases of an interval and of rotation-invariant domains,} see the examples in \cite{Grebenkov20c}
and in section \ref{sec:shell}). Note also that the expressions {\clr for the SCA} derived in
\cite{szabo,colloid,11,dist5,dist7,dist4,dist2} could be directly found from this
general solution for each considered geometric setting. An extension
of the expression \eqref{eq:Happ} to any $\x\in\Omega$ is simply
\begin{equation}
\label{eq:Happ2}
\tilde{H}_{\rm app}(p|\x)=\sum\limits_{n=0}^\infty\frac{V_n^{(p)}(\x)}{\mu_n^{(p)}
/J}C_n^*,
\end{equation}
with $V_n^{(p)}(\x)$ given by \eqref{eq:Vnp}.

The comparison of the spectral solution \eqref{eq:Hp_spectral} with \eqref{eq:Happ}
indicates that the SCA can formally be understood as a sort of approximate inversion
of the matrix $\Mu^{(p)}/q+\K^{(p)}$. More precisely, the SCA is retrieved if
\begin{equation}
\biggl[\bigl(\Mu^{(p)}/q+\K^{(p)}\bigr)^{-1}\biggr]_{n,n'}\approx\frac{\delta_{n,
n'}}{\mu_n^{(p)}/q}\left(1+\frac{q}{|\T|}\sum\limits_{n=0}^\infty\frac{|C_n|^2}{
\mu_n^{(p)}}\right)^{-1}.
\end{equation}
This specific form of approximate inversion is quite unexpected. The
first factor is the result of the inversion of the diagonal matrix
$\Mu^{(p)}/q$, if the matrix $\K^{(p)}$ was neglected. In turn, the
second factor, which does not depend on $n$, is a common correction to
all diagonal elements due to the matrix $\K^{(p)}$. It is worthwhile
noting that this approximation differs from the so-called diagonal
approximation developed in \cite{9,9a,9b,9c} in the context of
surface-mediated intermittent diffusion, in which the "correction"
matrix (like $\K^{(p)}$ here) was approximated by keeping only its
diagonal elements, allowing to obtain explicit approximate solutions.
We also stress that the limit $q\to \infty$ of a perfectly reactive target
is rather challenging for the exact spectral solution in \eqref{eq:Hp_spectral}
because the factor $1/q$ stands in front of the matrix $\Mu^{(p)}$
representing the dominant contribution of an unbounded operator $\M_p$,
as compared to the bounded operator $\I_{\T}$ represented by the
matrix $\K$.  Concurrently, this limit is trivial within the SCA.

In the next section we apply this general approach to spherical shell domains,
whose rotational symmetry greatly simplifies the above solutions and permits to
obtain explicit expressions.

\section{Spherical shell domains}
\label{sec:shell}

\begin{figure}
\begin{center}
\includegraphics[width=120mm]{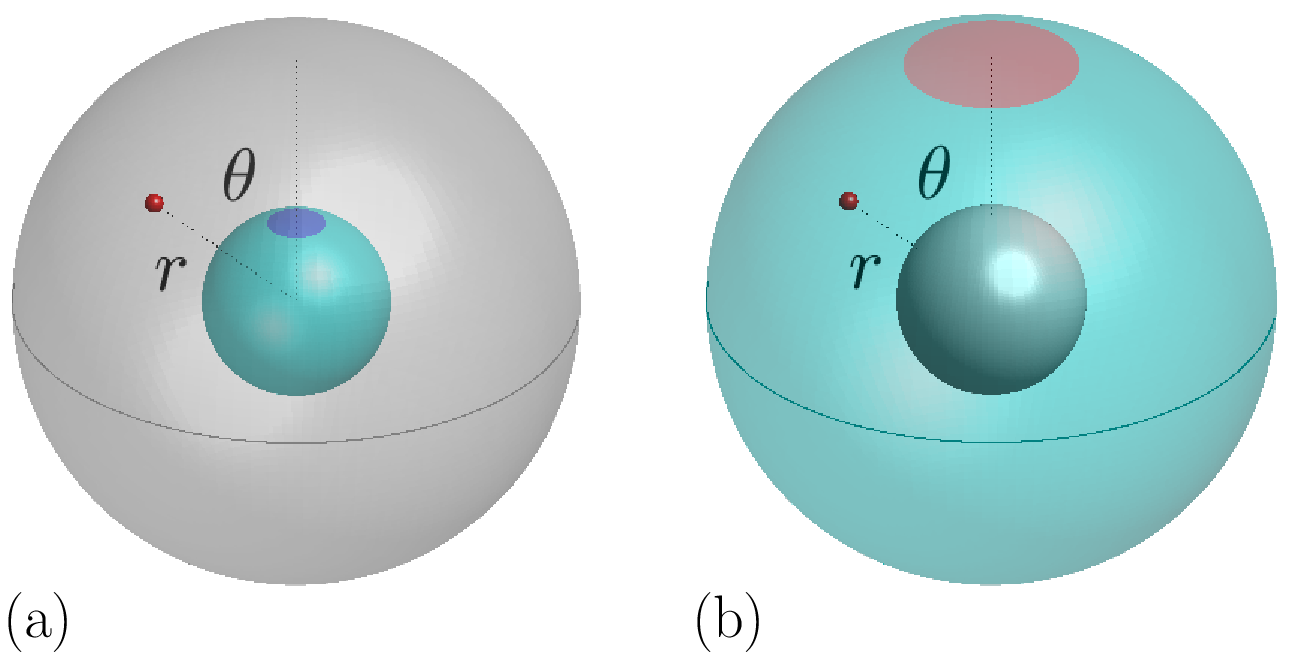}
\end{center}
\caption{Sketch of the geometrical setup in section \ref{sec:shell}: a spherical
shell {\clr domain} $\Omega$ is delimited by the impermeable boundaries of two concentric
balls, $\Omega_1$ with radius $R_1$ and $\Omega_2$ with radius $R_2$ ($R_1<R_2$).
The small red ball indicates the starting position of the diffusing particle,
with spherical coordinates $(r,\theta)$. Note that the azimuthal angle $\phi$
does not matter in such a geometry due to the axial symmetry. {\bf (a)} The
target (dark blue region) is located on the inner surface. {\bf (b)} The
target (light red region) is located on the outer surface.}
\label{fig:shell2}
\end{figure}

We consider two variants of the FRT problem in a spherical shell domain $\Omega
=\{\x\in\R^3~:~R_1<|\x|<R_2\}$, between two concentric spheres of radii $R_1<R_2$,
respectively (see figure \ref{fig:shell2}). The boundary $\pa$ of the domain is
fully reflecting except for a circular cap region at the North pole, the target,
defined in spherical coordinates $(r,\theta,\phi)$ by the inequality $\theta<\ve$.
This target can be located either on the boundary of the inner sphere (Problem
I) or on the boundary of the outer sphere (Problem II). In our further analysis we
concentrate on Problem I with the target on the boundary of the inner sphere. The
analysis for the Problem II is very similar and we will merely present the final
results without detailed derivation.

Since the boundary $\pa$ of the spherical shell domain consists of the boundaries
of two disjoint spheres, $\pa=\pa_1\cup\pa_2$, one of which is fully reflecting it
is convenient to modify the definition \eqref{eq:DtN} of the Dirichlet-to-Neumann
operator by restricting its action only on the sphere which contains the target
\cite{Grebenkov20c,Grebenkov20d}. For instance, for Problem I, one defines the
operator $\M_p$ that associates to a given function $f(\x)$ on $\pa_1$ another
function $g(\x)$ on the same boundary,
\begin{equation}
\M_p:f(\x)\to g(\x)=\left.\left(\frac{\partial u(\x)}{\partial n}\right)\right|
_{\pa_1},\mbox {where }\left\{\begin{array}{ll}(p-D\Delta)u(\x)=0 & (\x\in\Omega)\\
u(\x)=f(\x) & (\x\in\pa_1)\\
\partial u(\x)/\partial n=0 & (\x\in\pa_2)\end{array}\right..
\end{equation}

\subsection{Problem I. Target on the inner sphere}

We start with the exact spectral solution discussed in the previous
section. Since the Dirichlet-to-Neumann operator does not depend on the target
location, we take advantage of the rotational invariance of the shell domain
to determine the eigenbasis of $\M_p$ along with the matrices $\Mu^{(p)}$ and
$\K^{(p)}$ \cite{Grebenkov19a,Grebenkov20,Grebenkov20c}. Moreover, as the
circular target preserves the axial symmetry of the problem, the solution does
not depend on the angle $\phi$. One can therefore keep only the eigenfunctions
that are independent of $\phi$, see \cite{Grebenkov20c} for details,
\begin{equation}  \label{eq:vn_inner}
v_n(\theta)=\frac{1}{R_1}\sqrt{\frac{2n+1}{4\pi}}P_n(\cos\theta),
\end{equation}
where $P_n(z)$ are Legendre polynomials, and
\begin{equation}
\label{eq:mupI}
\mu_n^{(p)}=-g'_n(R_1),
\end{equation}
with
\begin{equation}
\label{eq:gnI}
g_n(r)=\frac{k'_n(\alpha R_2)i_n(\alpha r)-i'_n(\alpha R_2)k_n(\alpha r)}
{k'_n(\alpha R_2)i_n(\alpha R_1)-i'_n(\alpha R_2)k_n(\alpha R_1)}.
\end{equation}
Here $\alpha=\sqrt{p/D}$, $i_n(z)=\sqrt{\pi/(2z)}I_{n+1/2}(z)$ and $k_n(z)=
\sqrt{2/(\pi z)}K_{n+1/2}(z)$ are the modified spherical Bessel functions of
the first and second kind. The prime here and henceforth denotes the
derivative with respect to the argument. The radial functions $g_n(r)$ satisfy
the second-order differential equation
\begin{equation}  \label{eq:gn_diff}
g''_n(r)+\frac{2}{r}g'_n(r)-\frac{n(n+1)}{r^2}g_n(r)-\alpha^2 g_n(r)=0,
\end{equation}
with $g_n(R_1)=1$ and $g'_n(R_2)=0$. Note that here the eigenfunctions
$v_n (\theta)$ are independent of $p$.  Using the explicit form of the
Dirichlet Green's function from \cite{Grebenkov20c}, one also gets
$V_n^{(p)}(\x) = g_n(r) v_n(\theta)$.

One can also compute explicitly the matrix elements $\K_{n,n'}$ (see
equation (D12) of \cite{Grebenkov19a}), which are independent of the
Laplace parameter $p$,
\begin{equation}
\K_{n,n'}=\sqrt{(n+1/2)(n'+1/2)}\sum\limits_{k=0}^{\min\{n,n'\}} B_{nn'}^k
\frac{P_{n+n'-2k-1}(\cos\ve)-P_{n+n'-2k+1}(\cos\ve)}{2(n+n'-2k)+1},
\end{equation}
where 
\begin{equation}
B_{nn'}^k=\frac{A_k A_{n-k}A_{n'-k}}{A_{n+n'-k}}\frac{2n+2n'-4k+1}{2n+2n'-2k+1},
\qquad A_k=\frac{\Gamma(k+1/2)}{\sqrt{\pi}\Gamma(k+1)}.
\end{equation}
Note that a more general case of multiple non-overlapping circular targets was
also considered in \cite{Grebenkov19a}. Finally, the integral in \eqref{eq:Cn}
can be easily performed to yield
\begin{equation}
C_n=R_1\frac{\sqrt{\pi}}{\sqrt{2n+1}}\bigl(P_{n-1}(\cos\ve)-P_{n+1}(\cos\ve)\bigr).
\end{equation}
Substituting these expressions into \eqref{eq:Hp_spectral1} we get
\begin{equation}
\label{eq:Hp_spectral2}
\tilde{H}(p|\x)=\sum\limits_{n,n'=0}^\infty\sqrt{2n+1}g_n(r)P_n(\cos\theta)
\biggl[\bigl(\Mu^{(p)}/q+\K\bigr)^{-1}\biggr]_{n,n'}\frac{P_{n'-1}(\cos\ve)
-P_{n'+1}(\cos\ve)}{2\sqrt{2n'+1}},
\end{equation}
with the standard convention $P_{-1}(z)\equiv1$. 
Together with the explicit formulas for the matrices $\Mu^{( p)}$ and
$\K$, this is the exact solution of the original problem in
the Laplace domain. However, we are unable to invert the
infinite-dimensional matrix $\Mu^{(p)}/q+\K$ analytically and have to
resort to a numerical inversion.  In doing so we truncate the matrices
$\Mu^{(p)}$ and $\K$ at some order and exercise care afterwards that
finite-size effects do not matter, such that our numerical solution is
valid with any prescribed accuracy.

While our main focus is on the case of a fixed starting point
$\x$, it is instructive to consider two other common situations, in
which the starting point is uniformly distributed, either in the bulk,
or on a sphere of some radius $r$. The former case can be relevant
when the particle is produced inside the domain in a random location.
In turn, the second case accounts for situations when a cell has many
membrane channels for a given molecular species, and thus the release
point to the shell of interest on the inner membrane surface can be
viewed as randomly located. In both cases the above solution in
\eqref{eq:Hp_spectral2} is simplified. The surface-averaged solution reads
\begin{align}
\nonumber
\overline{\tilde{H}(p|r)}&=\frac{1}{4\pi r^2}\int\limits_{|\x|=r}d\x\tilde{H}(p|\x)\\
&=\sum\limits_{n'=0}^\infty g_0(r)\biggl[\bigl(\Mu^{(p)}/q+\K\bigr)^{-1}\biggr]_{0,
n'}\frac{P_{n'-1}(\cos\ve)-P_{n'+1}(\cos\ve)}{2\sqrt{2n'+1}},
\end{align}
where the orthogonality of the Legendre polynomials leads to the cancellation of
all terms with $n>0$. The volume-averaged solution involves an additional
integral over $r$,
\begin{align*} 
\overline{\tilde{H}(p)}&=\frac{1}{4\pi (R_2^3-R_1^3)/3}\int\limits_{\Omega}d\x
\tilde{H}(p|\x)=\frac{3}{R_2^3-R_1^3}\int\limits_{R_1}^{R_2}drr^2\overline{
\tilde{H}(p|r)}.
\end{align*}
Multiplying \eqref{eq:gn_diff} by $r^2$ and integrating over $r$ from
$R_1$ to $R_2$, one gets
\begin{equation}
\int\limits_{R_1}^{R_2}drr^2g_0(r)=\frac{R_2^2g'_0(R_2)-R_1^2g'_0(R_1)}{\alpha^2}=
\frac{R_1^2\mu_0^{(p)}}{p/D},
\end{equation}
from which
\begin{equation}
\overline{\tilde{H}(p)}=\frac{3DR_1^2\mu_0^{(p)}}{p(R_2^3-R_1^3)}\sum\limits_{n'=0}
^\infty\biggl[\bigl(\Mu^{(p)}/q+\K\bigr)^{-1}\biggr]_{0,n'}\frac{P_{n'-1}(\cos\ve)
-P_{n'+1}(\cos\ve)}{2\sqrt{2n'+1}}.
\end{equation}

We next turn to the SCA developed in the previous section. In the geometry
considered here our equation \eqref{eq:Happ2} becomes
\begin{equation}
\label{eq:Happ_inner}
\tilde{H}_{\rm app}(p|\x)=J\sum\limits_{n=0}^\infty g_n(r)P_n(\cos\theta)
\frac{1}{\mu_n^{(p)}}\frac{P_{n-1}(\cos\ve)-P_{n+1}(\cos\ve)}{2},
\end{equation}
where the parameter $J$ is determined from \eqref{eq:J} as
\begin{equation}
\label{eq:J_inner}
J=\left(\frac{1}{q}+\frac{1}{2(1-\cos\ve)}\sum\limits_{n=0}^\infty\frac{
\bigl(P_{n-1}(\cos\ve)-P_{n+1}(\cos\ve)\bigr)^2}{(2n+1)\mu_n^{(p)}}\right)^{-1},
\end{equation}
and we took into account the fact that the area $|\T|$ of the
spherical cap is given explicitly by $|\T|=2\pi R_1^2(1-\cos\ve)$.
The surface-averaged and volume-averaged approximations are
particularly simple,
\begin{equation}
\label{eq:Happ_surf}
\overline{\tilde{H}_{\rm app}(p|r)}=J\frac{g_0(r)}{\mu_0^{(p)}}\frac{1-\cos\ve}{2}
\end{equation}
and
\begin{equation}
\label{eq:Happ_vol}
\overline{\tilde{H}_{\rm app}(p)}=J\frac{3R_1^2D}{p(R_2^3-R_1^3)}\frac{1-\cos\ve}{2}
=J\frac{D|\T|}{p|\Omega|}.
\end{equation}
In \ref{sec:comparison}, we illustrate the remarkable agreement
between the prediction \eqref{eq:Happ_inner} of the SCA and the exact
solution \eqref{eq:Hp_spectral2}, even in the case when the target
covers half of the inner sphere.

Below we consider two particular cases in which the system reduces to
previously studied models.

\subsubsection*{The entire boundary of the inner sphere is a target.}

In the particular case when the target extends over the whole boundary
of the inner sphere, i.e., when $\ve=\pi$, the matrix $\K$ is the identity
matrix due to the orthonormality of the eigenfunctions $\{v_n^{(p)}\}$.
Moreover in this case all terms in \eqref{eq:Hp_spectral2} with $n>0$ vanish,
such that the exact spectral solution becomes
\begin{equation}
\label{eq:Hp_spectral_full}
\tilde{H}_{\ve=\pi}(p|\x)=\frac{g_0(r)}{1+\mu_0^{(p)}/q},
\end{equation}
where $g_0(r)$ and $\mu_0^{(p)}$ in \eqref{eq:mupI} and \eqref{eq:gnI} are
explicitly given by
\begin{align}
g_0(r)&=\frac{R_1}{r}\frac{e^{\alpha(R_1-r)}(e^{2\alpha r}(1+\alpha R_2)-e^{2
\alpha R_2}(1-\alpha R_2))}{e^{2\alpha R_1}(1+\alpha R_2)-e^{2\alpha R_2}(1-
\alpha R_2)},\\
\mu_0^{(p)}&=\frac{e^{2\alpha R_1}(1-\alpha R_1)(1+\alpha R_2)-e^{2\alpha R_2}
(1+\alpha R_1)(1-\alpha R_2)}{R_1\bigl(e^{2\alpha R_1}(1+\alpha R_2)-e^{2\alpha
R_2}(1-\alpha R_2)\bigr)} .
\end{align}
We thus retrieve the textbook solution for a spherical target surrounded by a
larger reflecting sphere (see, e.g., \cite{dist1} for a more detailed discussion).
Interestingly, the SCA result \eqref{eq:Happ_inner} also yields the exact result
\eqref{eq:Hp_spectral2} when $\ve\to\pi$. As a consequence, the SCA appears to be
accurate not only in the limit of a vanishingly small target,
$\ve\to0$, but also exact in the opposite limit $\ve\to\pi$. This is a
direct consequence of the self-consistent closure condition and one of
the reasons why the SCA yields accurate results even for the
intermediate case of large targets.

\subsubsection*{Limit $R_2\to\infty$.}

In the limit $R_2\to\infty$ the outer reflecting boundary extends to infinity
and one retrieves a common setting of a single circular target on a sphere,
explored by a particle diffusing in the unbounded space $\Omega=\{\x\in\R^3~:~
|\x|>R_1\}$. This is precisely the classical problem of binding of a ligand
that diffuses in an extracellular medium to a finite-sized receptor on an
impermeable boundary (see, e.g., \cite{berg,szabo}). Recall, however, that 
{\clr former studies} of this problem concentrated exclusively on the mean FRTs
(or mean reaction rates). Our analysis below shows that the full PDF of the
FRTs can be evaluated for such a geometrical setting.

Even though the domain itself is unbounded, its boundary $\pa$ is bounded, and the
above solution is still applicable. From the asymptotic behaviour of the modified
spherical Bessel functions $i_n(z)$ and $k_n(z)$ one finds that the radial functions
$g_n(r)$ converge to
\begin{equation}
g_n(r)\to-\frac{k_n(\alpha r)}{k_n(\alpha R_1)}.
\end{equation}
These determine the eigenvalues $\mu_n^{(p)}$ via equation \eqref{eq:mupI}. The
other quantities remain unchanged. To our knowledge, this is the first time
when exact and approximate expressions for the Laplace-transformed PDF $\tilde{H}
(p|\x)$ are presented in this geometric setting.

\subsection{Problem II. Target on the boundary of the outer sphere}

As we already remarked, the case of the target on the outer sphere is fairly
similar to the previously considered Problem I. Therefore, here we just list
the modifications by using the known form of the Dirichlet-to-Neumann
eigenbasis \cite{Grebenkov20c},
\begin{equation}
v_n(\theta)=\frac{1}{R_2}\sqrt{\frac{2n+1}{4\pi}}P_n(\cos\theta)
\end{equation}
and
\begin{equation}
\label{eq:mup}
\mu_n^{(p)}=g'_n(R_2),
\end{equation}
with
\begin{equation}
g_n(r)=\frac{k'_n(\alpha R_1)i_n(\alpha r)-i'_n(\alpha R_1)k_n(\alpha r)}
{k'_n(\alpha R_1)i_n(\alpha R_2)-i'_n(\alpha R_1)k_n(\alpha R_2)},
\end{equation}
satisfying $g_n(R_2)=1$ and $g'_n(R_1)=0$.\footnote{Note that there is a
misprint in equation (B2) of \cite{Grebenkov20c}, in which $g_n'(R)$ should
be replaced by $g_n'(L)$.} In turn, the matrix $\K$ remains unchanged, as
well as the formulae \eqref{eq:Hp_spectral2}, \eqref{eq:Happ_inner}, and
\eqref{eq:J_inner} for $\tilde{H}(p|\x)$, $\tilde{H}_{\rm app}(p|\x)$, and $J$.

In the limit $R_1 \to 0$, the reflecting inner sphere shrinks to a point, and
hence, does not hinder anymore the particle dynamics. One thus retrieves the
NEP for an escape from a sphere through a circular hole on its boundary. In
this limit, radial functions $g_n(r)$ converge to
\begin{equation}
g_n(r)\to\frac{i_n(\alpha r)}{i_n(\alpha R_2)},
\end{equation}
and we therefore recover the SCA result derived in \cite{dist2}
[indeed, equation (A.21) from \cite{dist2}, together with equations
(A.13, A.17, A.18), is identical with equations \eqref{eq:Happ_inner},
\eqref{eq:J_inner}].  The remarkable accuracy of the SCA prediction is
illustrated in \ref{sec:comparison}.

\subsection{The PDF in time domain}

The solution $H(t|\x)$ in time domain can be obtained via inverse Laplace
transformation by using the residue theorem. As the spherical shell domain
is bounded, the Laplace operator with mixed boundary conditions governing
the diffusion-reaction dynamics in equations \eqref{eq:diffusion} to
\eqref{eq:boundary} has a discrete spectrum, i.e., $\tilde{H}(p|\x)$ has
infinitely many poles $\{ p_k\}$ lying on the negative axis in the complex
plane $p\in \C$. Formally, these poles are determined as zeros of the
determinant of the matrix $\Mu^{(p)}/q+\K^{(p)}$ when the matrix is not
invertible, yielding a singular, pole-like behaviour of $\tilde{H}(p|\x)$.
If $\{\gamma_n^{(p)}\}$ are the eigenvalues of this matrix, then the
poles correspond to such values of $p$ at which at least one $\gamma_n^{
(p)}$ is zero. The poles determine the eigenvalues of the Laplace operator:
$\lambda_n=-p_n/D$. Evaluating the residues of $\tilde{H}(p|\x)$, one
formally gets the spectral expansion of the solution in time domain:
\begin{equation}
H(t|\x)=\sum\limits_{n=1}^\infty c_n(\x)e^{-Dt\lambda_n},
\end{equation}
where the $c_n(\x)$ are determined by these residues.

In practice it is more convenient to search the poles $\{ p'_k\}$ of $\tilde{
H}_{\rm app}(p|\x)$, that can be used as approximations of $\{p_k\}$. The
approximate poles are determined by an explicit equation on $p$, at which
the parameter $J$ from \eqref{eq:J_inner} diverges,
\begin{equation}
\label{mus}
\frac{1}{2(1-\cos\ve)}\sum\limits_{n=0}^\infty\frac{\bigl(P_{n-1}(\cos\ve)
-P_{n+1}(\cos\ve)\bigr)^2}{(2n+1)\mu_n^{(p)}}=-\frac{1}{q}.
\end{equation}
The computation of the residue at each pole $p'_k$ and their properties are
discussed for the case of a sphere in \cite{dist2}. Instead of this analysis
we will invert the Laplace transform numerically by using the Talbot algorithm.

\begin{figure}
\begin{center}
\includegraphics[width=70mm]{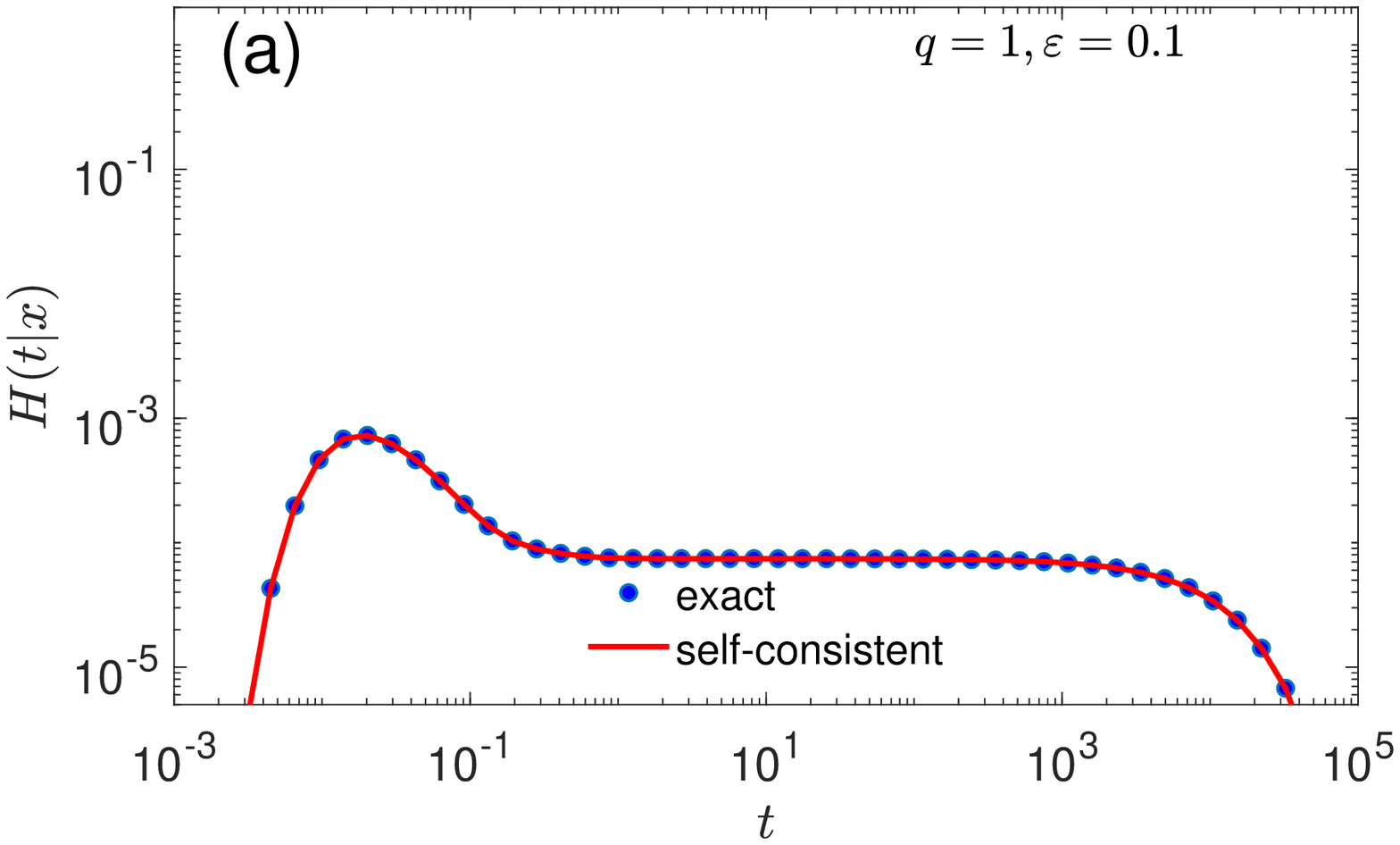}
\includegraphics[width=70mm]{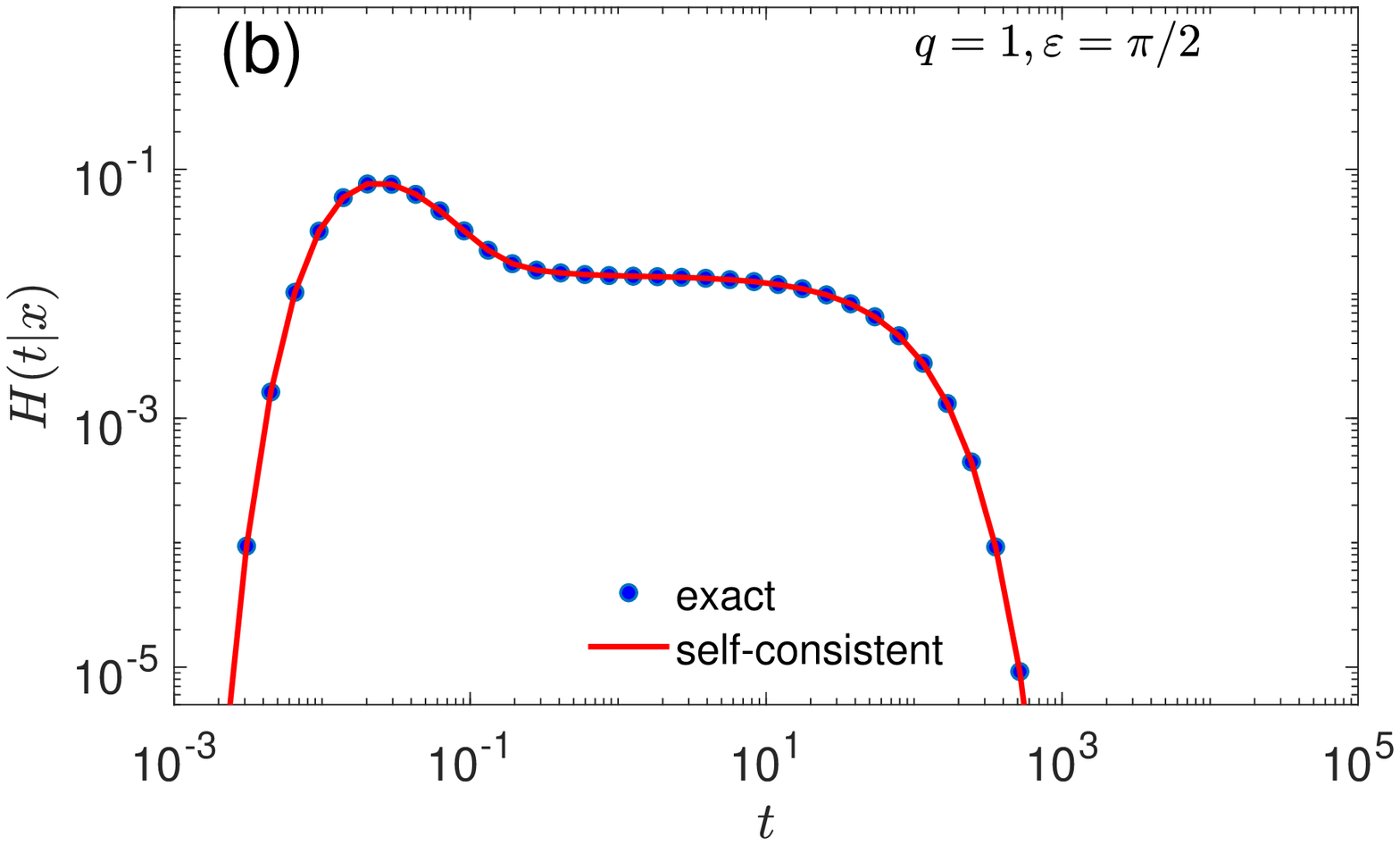}
\includegraphics[width=70mm]{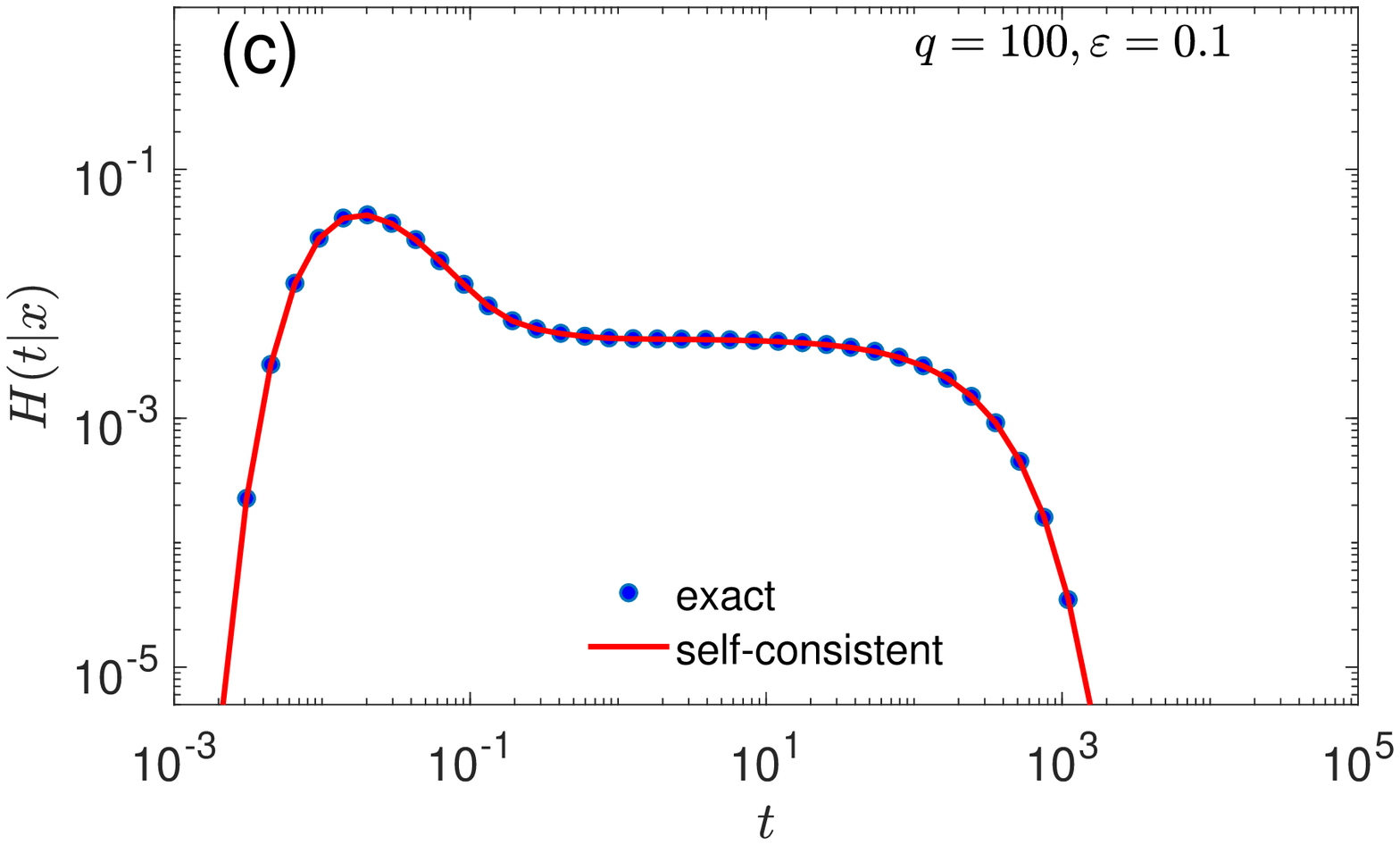}
\includegraphics[width=70mm]{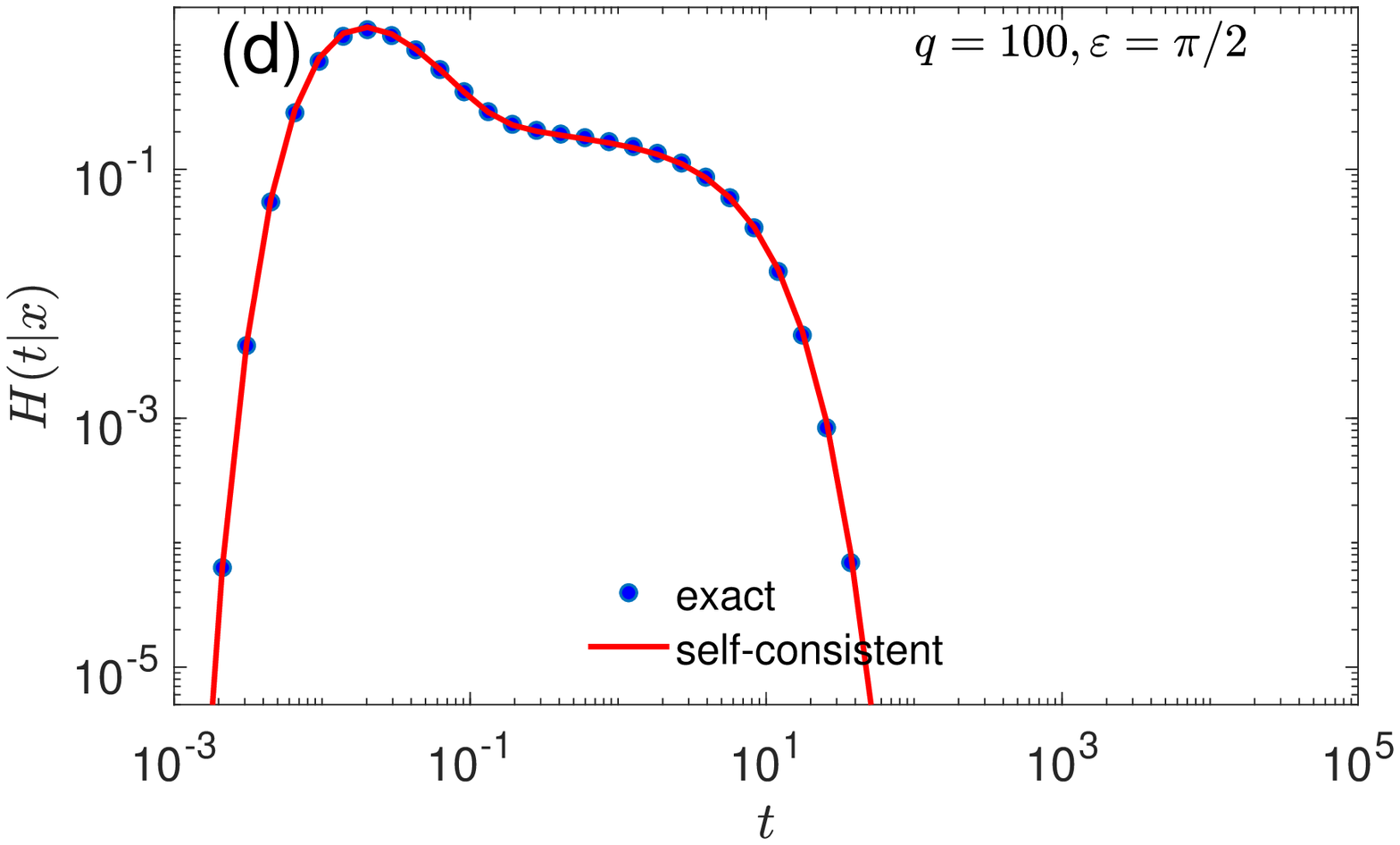}
\end{center}
\caption{Comparison between the exact solution $H(t|\x)$ (filled circles) and the
SCA result $H_{\rm app}(t|\x)$ (solid line) both of which are obtained
numerically via inverse Laplace transformation of $\tilde{H}(p|\x)$
and $\tilde{H}_{\rm app}(p|\x)$ by using the Talbot algorithm. The
latter quantities were calculated with truncation at $n_{\rm
max}=25$. The target is located at the inner sphere, with parameters:
$R_1=0.1$, $R_2=1$, $r=0.45$, $\theta = 0$, and $D=1$. {\bf (a)}
$q=1$, $\ve=0.1$; {\bf (b)} $q=1$, $\ve=\pi/2$; {\bf (c)} $q=100$,
$\ve=0.1$; {\bf (d)} $q=100$, $\ve=\pi/2$.}
\label{fig:Ht_inner}
\end{figure}

\begin{figure}
\begin{center}
\includegraphics[width=70mm]{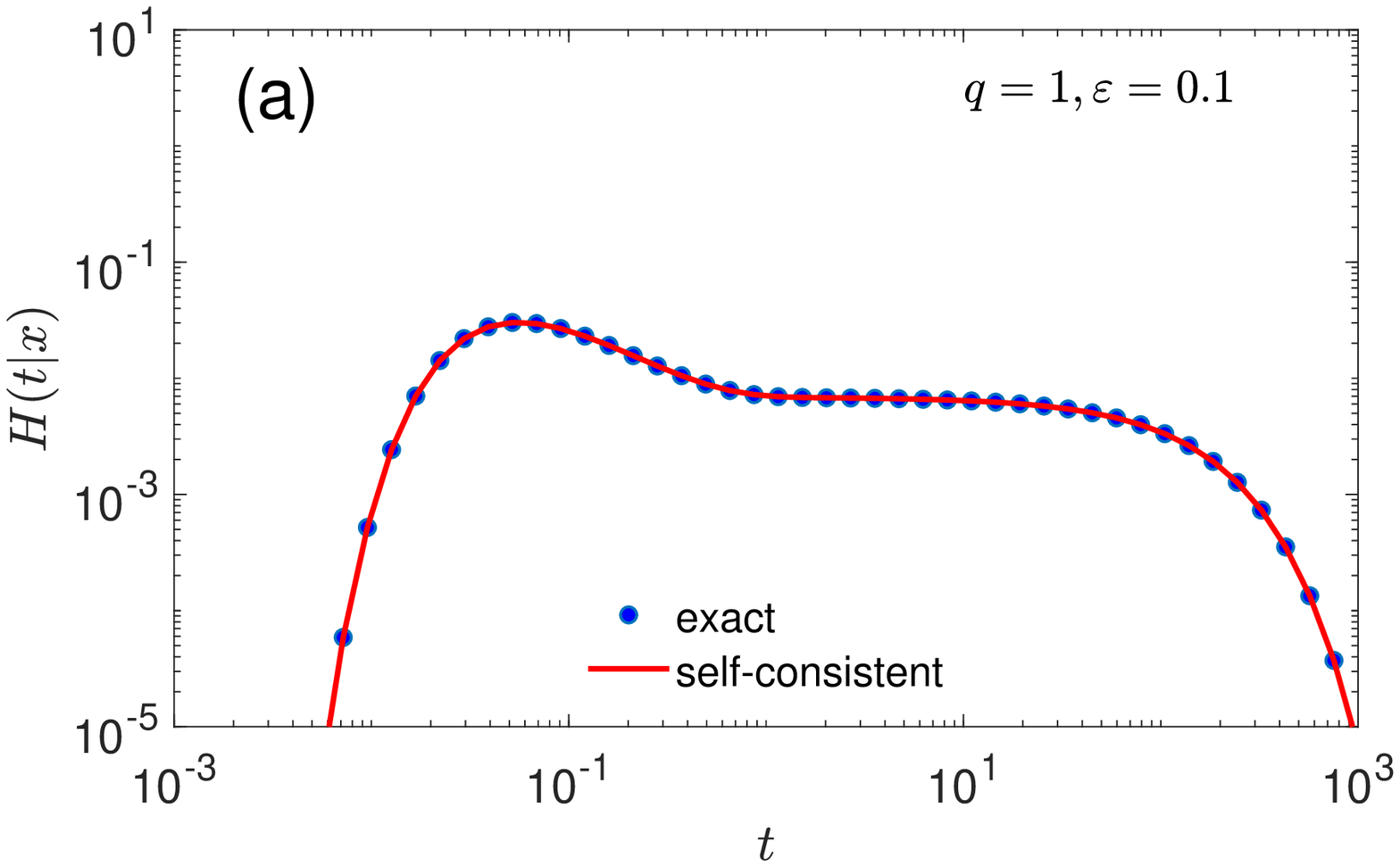}
\includegraphics[width=70mm]{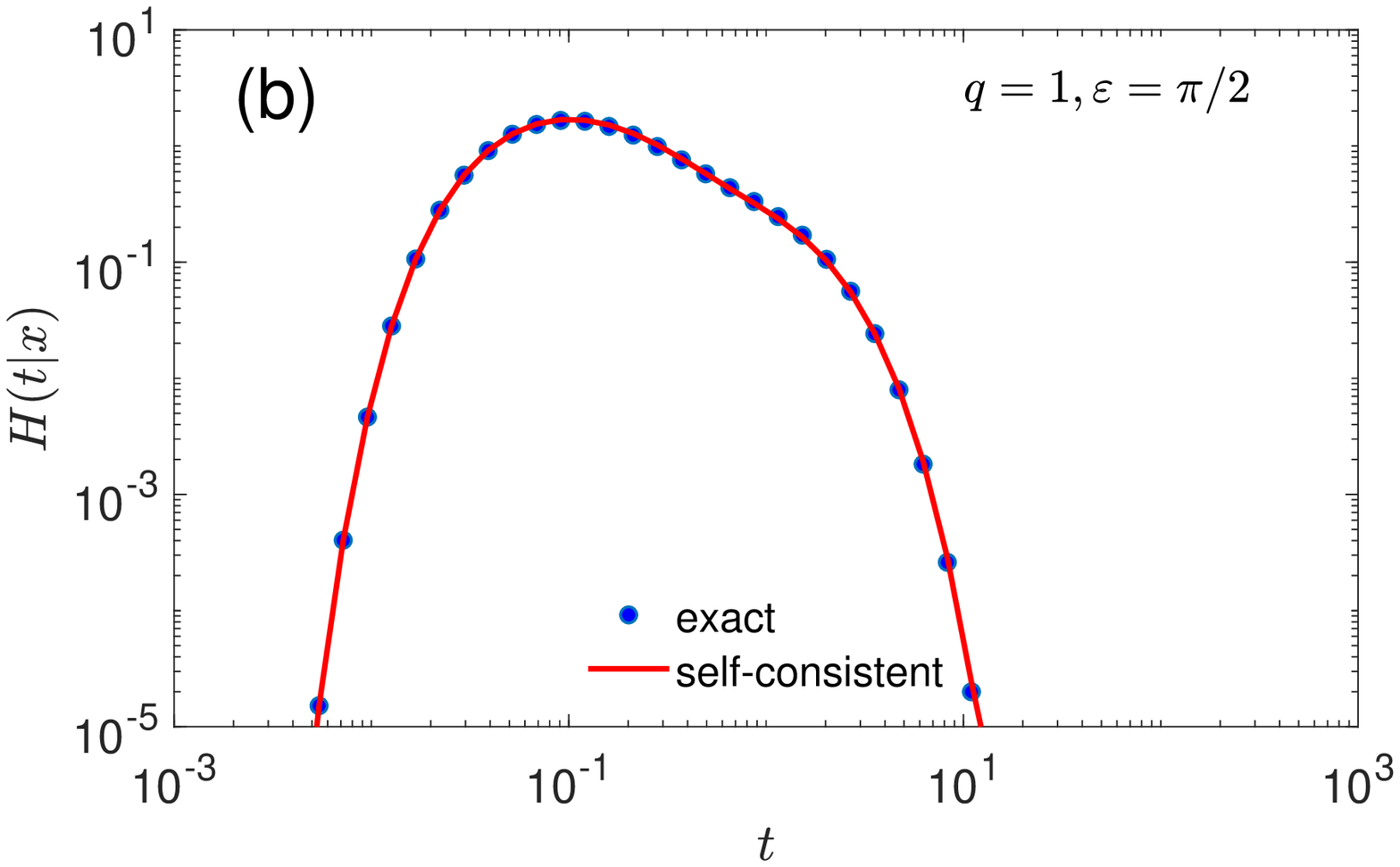}
\includegraphics[width=70mm]{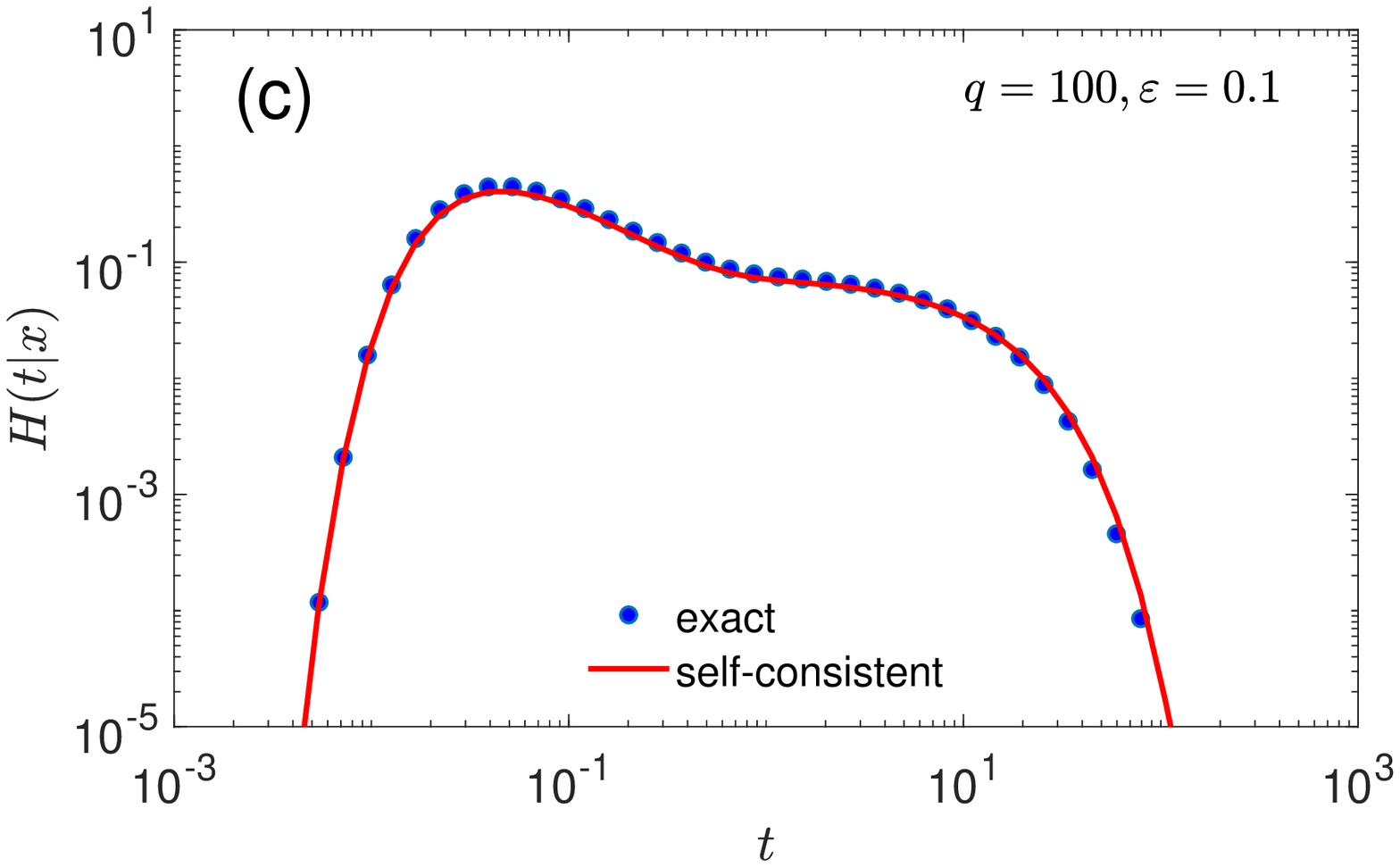}
\includegraphics[width=70mm]{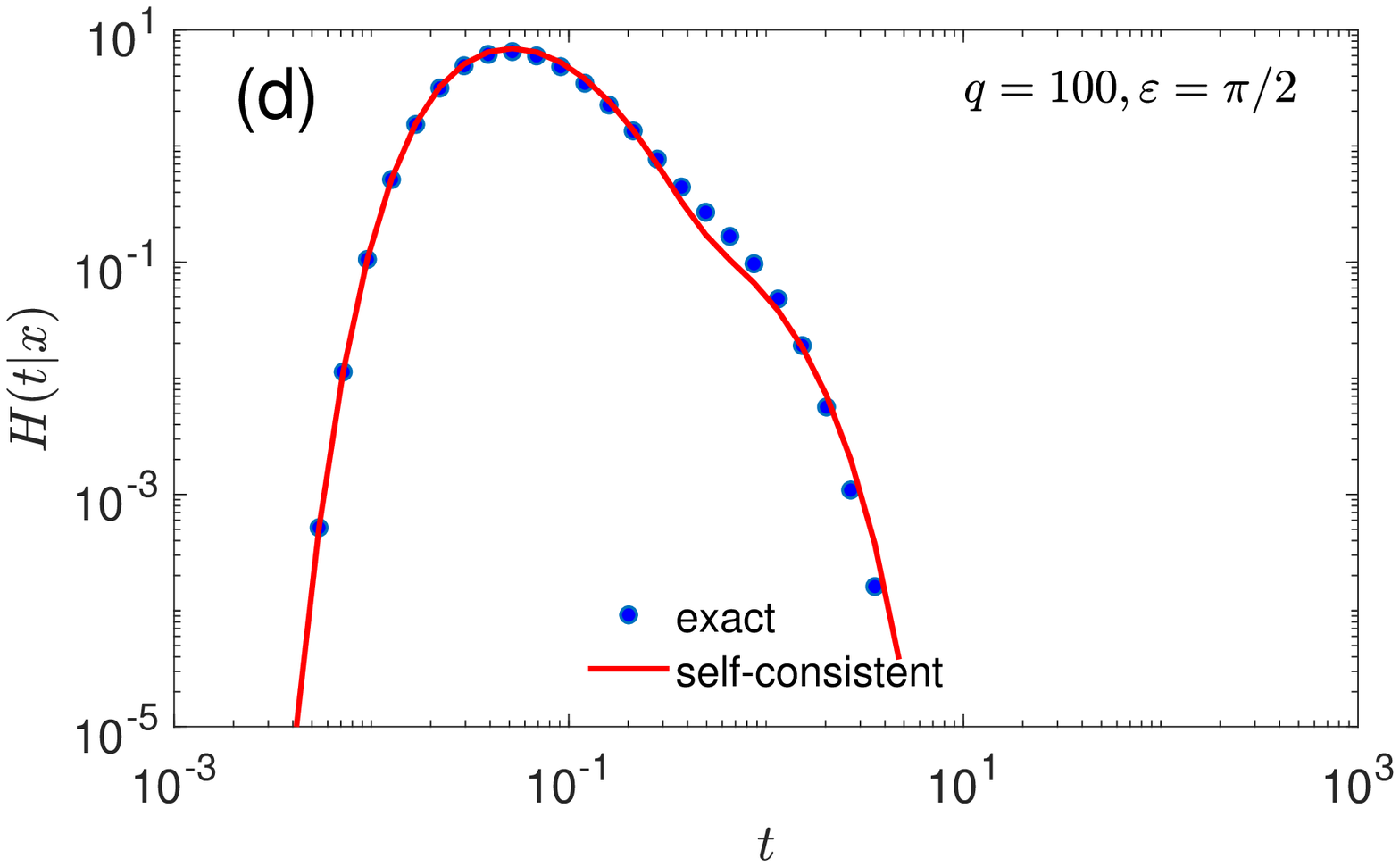}
\end{center}
\caption{
Comparison between the exact spectral solution for $H(t|\x)$ (filled
circles) and the SCA result $H_{\rm app}(t|\x)$ (solid line) both of
which are obtained numerically via the inverse Laplace transform of
$\tilde{H}(p|\x)$ and $\tilde{H}_{\rm app}(p|\x)$ by help of the
Talbot algorithm. The latter quantities were calculated with
truncation at $n_{\rm max}=25$. The target is located at the outer
sphere, with parameters: $R_1=0.1$, $R_2=1$, $r=0.45$, $\theta = 0$,
and $D=1$. {\bf (a)} $q=1$, $\ve=0.1$; {\bf (b)} $q=1$, $\ve= \pi/2$;
{\bf (c)} $q=100$, $\ve=0.1$; {\bf (d)} $q=100$, $\ve=\pi/2$.}
\label{fig:Ht_outer}
\end{figure}

Figure \ref{fig:Ht_inner} illustrates the behaviour of the PDF
$H(t|\x)$ and its self-consistent approximation $H_{\rm app}(t|\x)$
for a small target located on the inner sphere. 
Expectedly, the PDF is broader when the target is smaller and less
reactive. {\clr In particular, for the case $q = 1$ and $\ve = 0.1$,
the PDF spans over 8 orders of magnitude in time.}  Figure
\ref{fig:Ht_outer} presents the same quantities for the target located
at the outer sphere.  {\clr While we kept the same angular sizes of
the target, their linear sizes are $R_2/R_1 = 10$ times larger
here, which explains why the distributions are narrower in figure
\ref{fig:Ht_outer} in comparison to figure \ref{fig:Ht_inner}.}
We discuss these PDFs in the next section.

\section{Discussion}
\label{sec:discussion}

In this section we discuss several features of the calculated PDF $H(t|\x)$
such as the different regimes it exhibits as well as the asymptotic behaviour.
We also evoke the question of the "screening" effects of the boundary on, e.g.,
a perfectly-reactive target ($\kappa=\infty$) and finally analyse the
"concealment" effect of the inner sphere.

\subsection{The structure of the PDF in time domain}

One may notice that the PDF $H(t|\x)$ evaluated in the previous section 
as well as the exact spectral solution 
exhibits four different temporal regimes
delimited by three characteristic time scales. Namely, an initial regime
that extends from time zero up to the most probable (typical) FRT $t_{\mathrm{
mp}}$; a second regime in which the PDF descends from the maximal value; then
a plateau-like regime and, ultimately, an exponential decay at long times. We
note that this is quite a generic behaviour of the PDFs of the FRTs in
bounded domains, which has been amply discussed in previous papers, see, e.g.,
\cite{dist2,dist4,dist1}. Here, for the sake of completeness, we present a
succinct account of the different temporal regimes and the corresponding
asymptotic behaviour.

At short times $H(t|\x)$ is characterised by the singular L\'evy-Smirnov-type
form \cite{redner}
\begin{align}
\label{LS}
H(t|\x)\sim A_t\frac{\exp\left(-\rho^2/(4Dt)\right)}{t^{3/2}}\quad(t\to0),
\end{align} 
where $A_t$ is a computable amplitude and $\rho$ is the shortest distance
between the starting point and the target. This regime is thus dominated
by the "direct" trajectories \cite{aljaz,aljazprx,dist1}, which go straight from the
initial point to the target. As a consequence, in this regime the
dimensionality of the embedding space and of the boundaries does not come
into play, unless a particle starts close to the South pole while the target
is located at the North pole, such that the optimal path is "interacting"
with either of the boundaries. Note that $A_t$ in \eqref{LS} is $t$-independent
for perfect target {\clr ($\kappa = \infty$),} and is an algebraic
function of time, $A_t\sim \kappa t$, for imperfect reactions. In the latter
case the particle may be reflected from the target and react with it only
upon some subsequent arrival to its location. Therefore, in this latter case
$H(t|\x)$ is smaller for short times than in the case of perfect reactions.

The second regime corresponds to a power-law descent from the maximum. This
regime is also universal, in the sense that it is independent of the actual
dimensionality of space, and it terminates at the characteristic time $t_c$
when a particle first engages with any of the boundaries; thus realising
that it moves in a bounded domain. Note that this intermediate power-law
decay is responsible for an effective broadness of the PDF. As demonstrated
earlier in \cite{carlos1,carlos2}, in situations when this regime lasts
sufficiently long, the FRTs observed for two realisations of the process may
become disproportionally different.

Next a plateau-like regime with a very slow variation of $H(t|\x)$ emerges
due to the gap between the first and second eigenvalues $\lambda_1$ and
$\lambda_2$ (which are related to the poles, i.e., the solutions of equation
\eqref{mus} above). In this regime all values of the FRT are nearly equally
probable.

Finally, at times $t\sim1/(D\lambda_1)$ the PDF crosses over to an
exponential decay of the form
\begin{align}
H(t|\x)\sim\exp\left(-Dt\lambda_1\right).
\end{align}

{\clr In summary, there is} an appreciable amount of the trajectories that lead to much earlier
reaction times, see the discussions in \cite{aljaz,aljaz1,dist1,dist4,dist2}. In
the case of very small target, the \textit{mean\/} FRT is close to $1/(D
\lambda_1)$ (see \ref{sec:MFPT}), and thus the mean FRT {\clr controls the long-time behaviour of the PDF \cite{Benichou10f}.  
However, the mean FRT, which is typically orders of magnitude longer
than the typical reaction time $t_{\rm mp}$, is therefore insufficient
to describe the rich structure of the PDF, especially in the limit of
low concentrations typical for many biochemical situations
\cite{Reva21}.  Strong defocusing of the FRTs is a generic feature of
diffusion-controlled reactions, which was commonly ignored in former
studies.}

\subsection{Screening effects of the boundary}

As we have already remarked, the main mathematical difficulty of the problem
at hand is represented by the mixed boundary conditions: a zero-flux
boundary condition imposed on the reflecting parts of the boundary and a
reactive condition at the target. From a physical point of view, this
also implies that the reflecting part of the surface has its effect on the
reaction and may, to some unknown extent, "screen" the target.  To get
some (at least partial) understanding of an emerging effective screening, we
compare here the PDFs calculated for two {\clr cases:  the particle starts at the North
Pole of the outer sphere and searches for the target region at the North Pole
of the inner sphere (Problem I), and the particle starts
at the North pole of the inner sphere and searches for the target region at
the North Pole of the outer sphere (Problem II).}  The initial distance here is the same by definition, the
{\clr domain} in which a particle moves is the same, but in one case the surface in
the vicinity of the target is concave, while in the other it is convex.
There is, however, a small subtlety concerning the size of the target:
to render the two cases identical, should one take the same area of the
target or the same angular size? We consider both situations.

\begin{figure}
\begin{center}
\includegraphics[width=70mm]{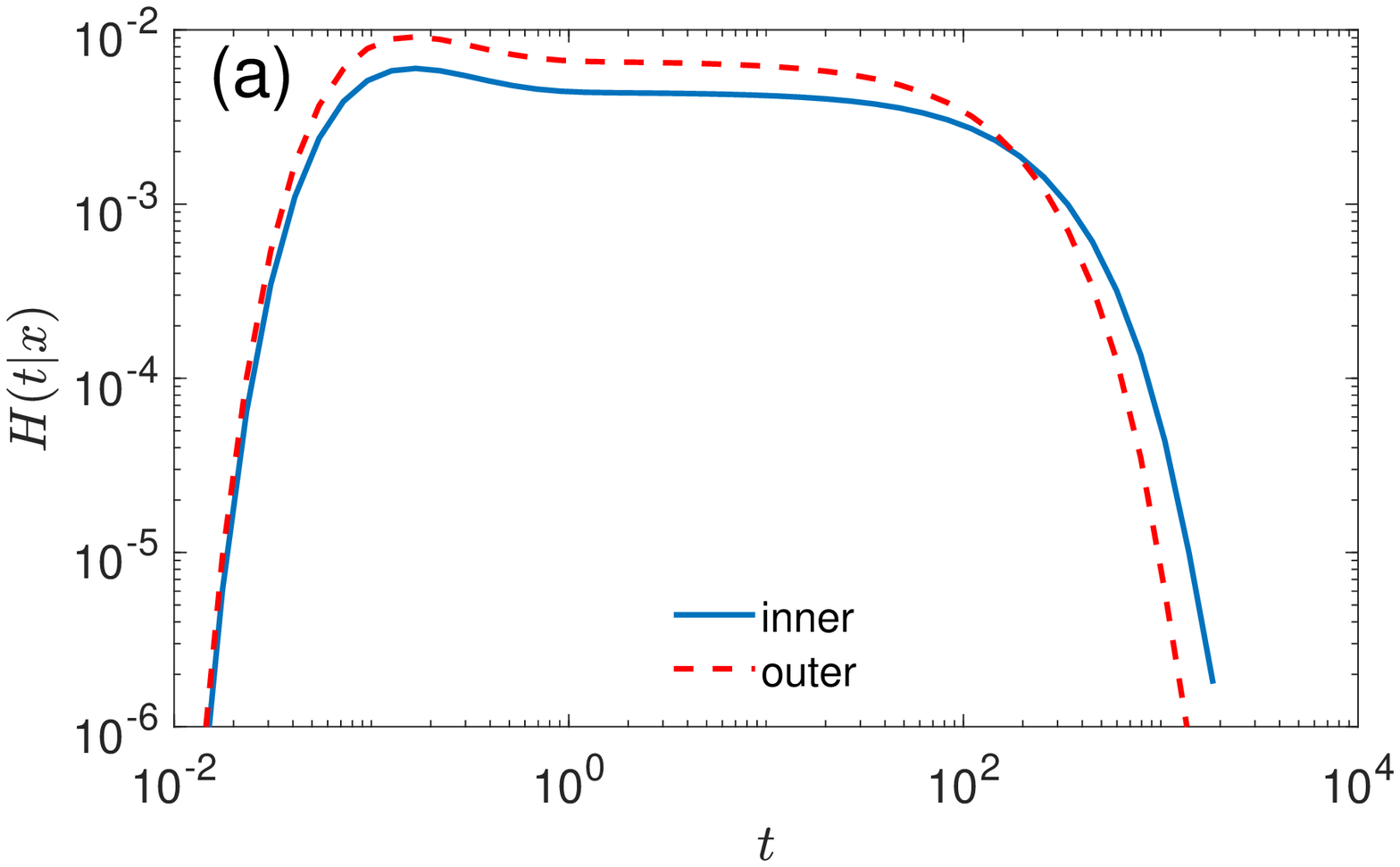}
\includegraphics[width=70mm]{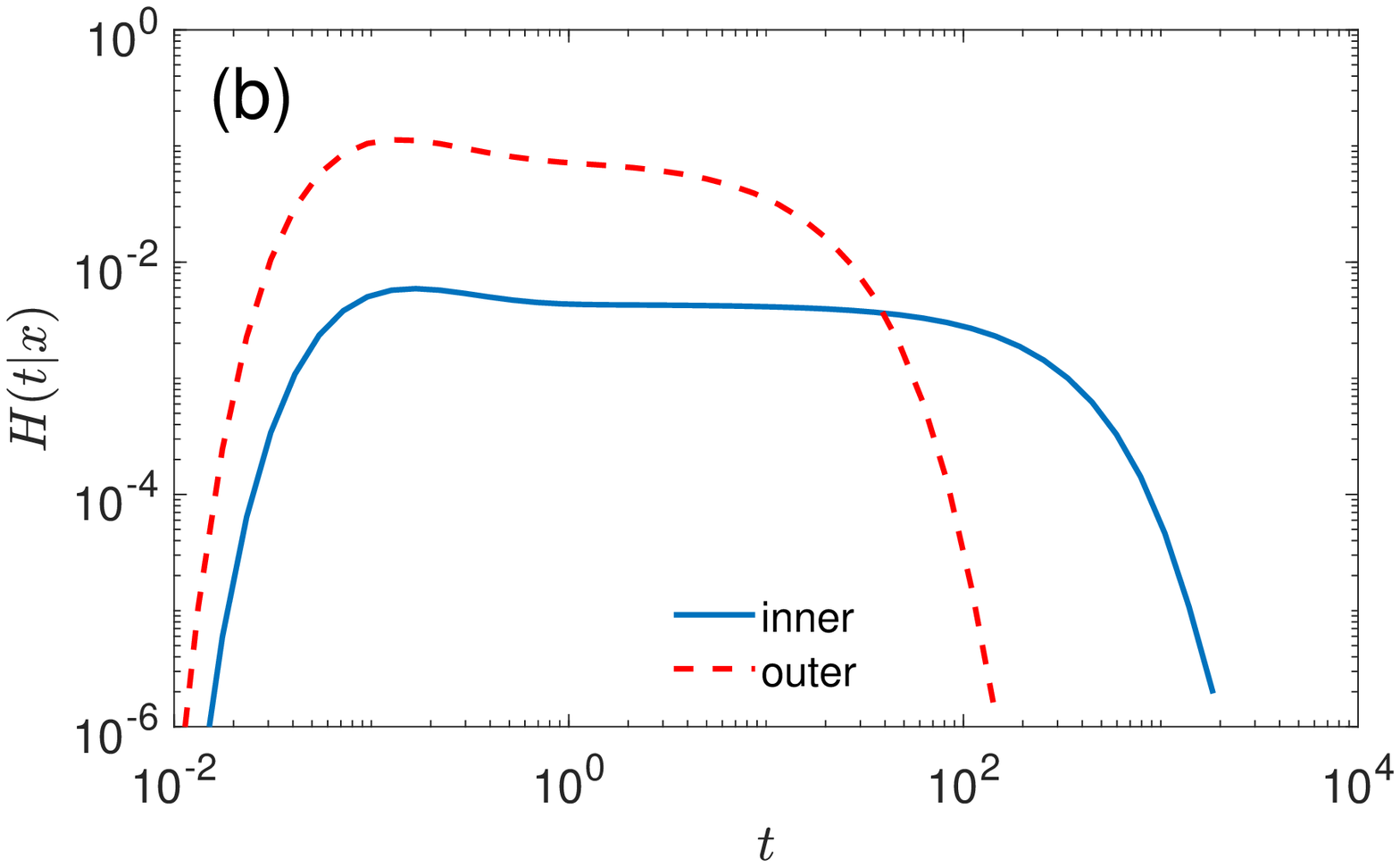}
\end{center}
\caption{\clr 
Comparison of the FRT PDFs $H(t|\x)$ for outer-to-inner case
(Problem I, when the particle starts from the North Pole of the outer
sphere ($r = R_2,~\theta =0$) and the target is at the North pole of
the inner sphere; solid blue curve) and inner-to-outer case (Problem
II, when the particle starts from the North pole of the inner sphere
($r = R_1,~\theta = 0$) and the target is at the North Pole of the
outer sphere; red dashed curve).  Parameters: $q=100$, $D=1$,
$R_1=0.1$, and $R_2=1$. {\bf (a)}: Areas of the target are
equal for problems I and II. The angular size of the target is
$\ve_1=0.1$ for I and $\ve_2=0.01$ for II. {\bf (b)}: Angular size of
the target site $\ve=0.1$ is the same for problems I and II.}
\label{fig:Ht_comp}
\end{figure}

In figure \ref{fig:Ht_comp} we compare the FRT PDFs $H(t|\x)$ for situations
(i) and (ii) in the case of a high intrinsic reactivity, $q=100$, which permits
us to disentangle the effects of a finite reaction probability from the screening
effects of the geometry. Clearly, for such a high reactivity, a particle most
likely reacts with the target upon first encounter such that $H(t|\x)$ should
be very close to the first-passage time PDF. We observe that, indeed, a
curvature in the vicinity of the target does matter, regardless of whether
we fix the area of the target or its angular size. 
When the area is fixed (figure \ref{fig:Ht_comp}(a)), the PDF is {\clr
slightly broader for the outer-to-inner case (Problem I) than for
the inner-to-outer case (Problem II),} and also the height of the
maximum is {\clr lower}.  {\clr As a consequence, transmitting signals
from the membrane to the nucleus, under similar biophysical
parameters, should usually take longer than in the opposite direction. This
seems highly relevant to cell signalling in general, and possibly provides
another rationale for the presence of concentric cytoskeletal tracks
in the cells.\footnote{This effect is related to the "centrifugal drift" or
"geometric spurious drift" mentioned in \cite{redner,aljazprx}.}}
In the case when $\ve$ is identical (figure
\ref{fig:Ht_comp}(b)), the effect is more pronounced and the
difference is more appealing: the distribution is {\clr much broader}
and the maximum is an order of maximum {\clr lower} in {\clr the
outer-to-inner case than in the inner-to-outer case.}  Therefore, the
FRTs in the case when the target is concave are more
focused around the most probable value.  We note parenthetically that
the effect of a local curvature on the first-passage times was already
studied within the context of the NEP. In particular, the mean
first-passage time was discussed in \cite{24,25,26} for the case when
the escape window is located at a corner, or at a cusp in the boundary
and on a Riemannian manifold. It was shown there that the very
functional dependence of the mean first-passage time on the angular
size $\ve$ of the aperture can be different, as compared to the case
when the boundary is smooth. Our results in figure \ref{fig:Ht_comp}
demonstrate that even in the case of a smooth boundary the effect of
the local curvature is present.

\subsection{Concealment effect of the inner sphere}

We turn here to the situation in which the target is located at the North
pole of the outer sphere, whereas the starting point is on the South pole
of the outer sphere. Evidently, this model can be viewed as the NEP in which
the particle dynamics is concealed by the impermeable inner sphere {\clr (an obstacle).} However,
the concealment effect is not evident \textit{a priori}: On the one hand, one
may expect that the presence of such an obstacle should slow down the search
process, because the particle now cannot move along "direct trajectories"
from the South to the North pole of the outer sphere and has to bypass the
obstacle, which makes its trajectories  somewhat longer.  On the other hand,
when the radius of the inner sphere becomes comparable to that of the outer
one (i.e., $R_1$ is close to $R_2$), the particle diffuses in a thin
spherical shell and thus undergoes effectively two-dimensional diffusion
that may speed up the search process. Which argument is valid here?  For a
different geometric setting \cite{Rupprecht17}, it was argued that obstacles
cannot speed up the search process. We thus pose the question of the effect
that the radius $R_1$ of the inner sphere has on the shape of the PDF.

\begin{figure}
\begin{center}
\includegraphics[width=70mm]{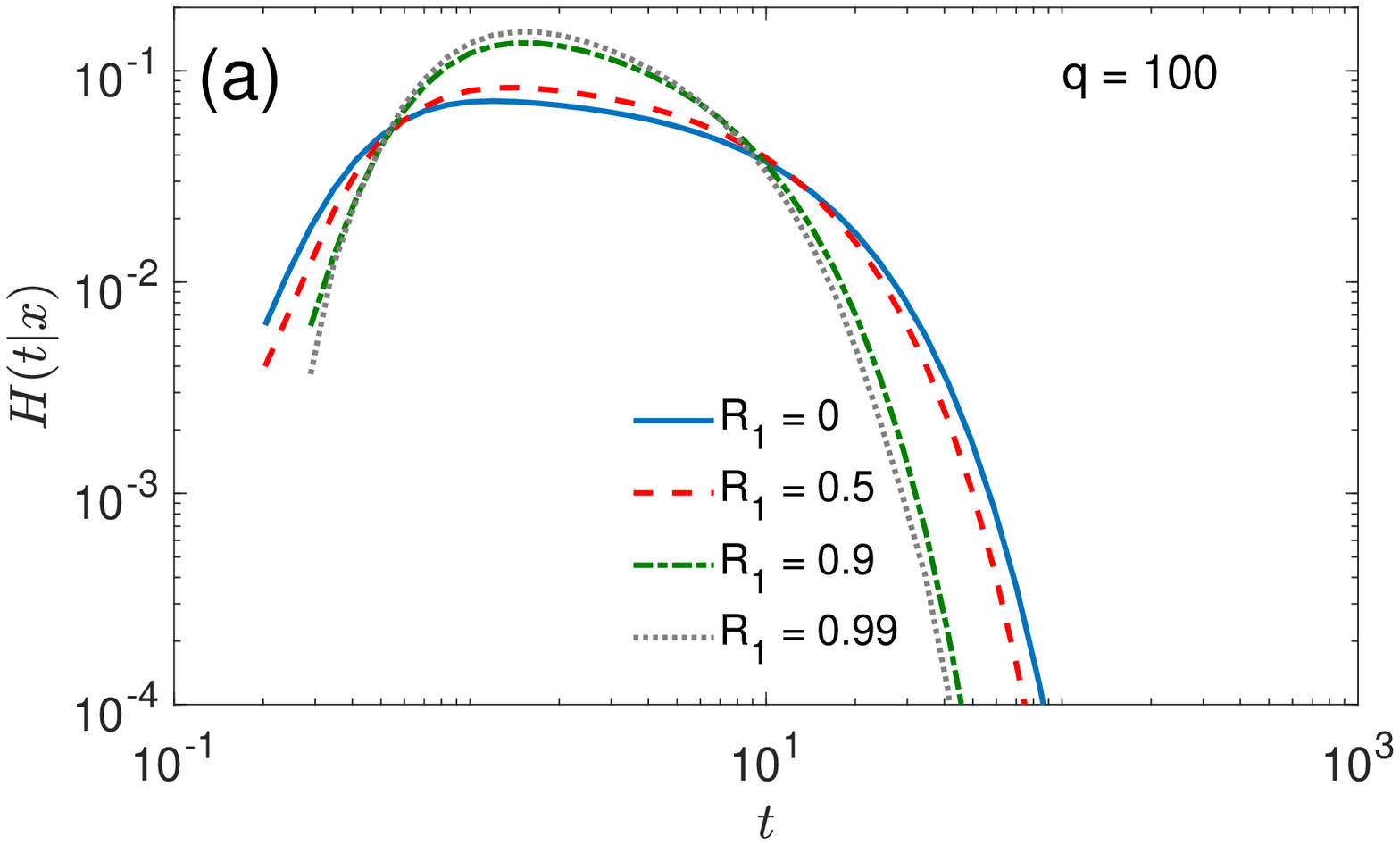}
\includegraphics[width=70mm]{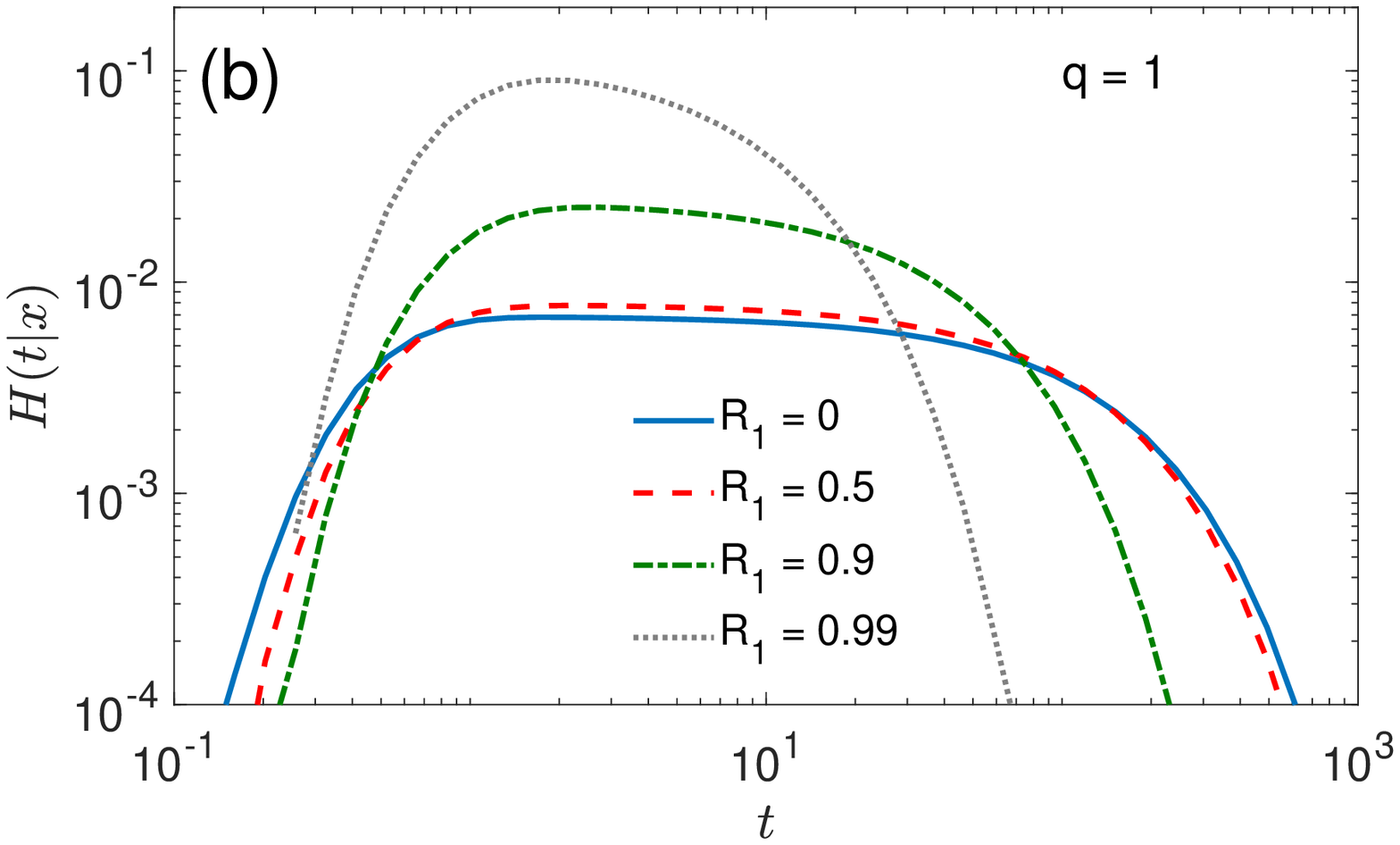}
\end{center}
\caption{FRT PDF to a target on the outer sphere, with $R_2=1$, $D=1$, $\ve=0.1$,
$q=100$ {\bf (a)}, and $q=1$ {\bf (b)}, the starting point at the
South pole of the outer sphere ($r=R_2$, $\theta=\pi$), and four
values of $R_1$. Note that some curves are not shown for too
short times due to numerical artifacts of the Laplace transform
inversion.}
\label{fig:Ht_obstacle}
\end{figure}

In figure \ref{fig:Ht_obstacle} we depict $H(t|\x)$ for the particular
case when the starting point of the particle is located at the
South pole of the outer sphere, the radius of the outer sphere is
fixed, and the radius of the inner sphere is varied. We find that, in
general, the PDF gets more focused around the most probable FRT 
upon an increase of the radius of the inner sphere, 
the peak value of the distribution at this point increases, and 
the distribution becomes more narrow, which implies that the variability of
the FRT reduces in the presence of an obstacle.

\section{Conclusions and perspectives}
\label{sec:conclusion}

Experimental progress in monitoring biochemical reactions has been massive in
the last two decades. Thus it is now possible to monitor gene expression events
such as real-time protein production "one protein molecule at a time" \cite{yu},
providing unprecedented insight into, e.g., bursty protein production \cite{yu},
specific binding of transcription factor proteins to their binding site on the
DNA \cite{elf}, or elucidating the molecular origins of bacterial individuality
\cite{wang}. Similarly, single-molecule biochemical reactions in the membranes
of living cells can be monitored and rationalised \cite{weigel}. The control of
gene expression is often running off at extremely small concentrations of the
involved molecules. Thus, the well-studied Lac and phage lambda repressor
proteins occur at copy numbers of only few to few tens in a single, relatively
small bacteria cell, corresponding to nanomolar concentrations. It was found that
even in such bacteria cells the distance between a gene encoding for a protein A
and another gene, to which A is supposed to bind to controls its expression, is
relevant \cite{kepes,kolesov}. Therefore it was concluded that a single reaction
rate, as used
since Smoluchowski's seminal work \cite{smoluchowski}, cannot account for the
complexity of the involved diffusive search \cite{otto} and biochemical reactions
themselves \cite{otto1}. Concurrently the role of heterogeneity in cells is being
recognised as a key element in understanding intracellular transport
\cite{kuehn,han,ma,aljaz,Doris}.

Here we studied the full statistics of first-reaction times to an imperfect
immobile target in a typical setting for intracellular reactions \cite{ma}.
In our model a particle starts from a fixed location and diffuses in a bounded
region between two nested impermeable domains, and the target is assumed
to be placed on either boundary, a typical situation for many molecular
regulation processes naturally running off in biological cells. As in previous
work \cite{aljaz,dist1,dist2,dist4,race,N5} we obtained clear evidence that the
associated reaction times have a broad distribution and thus the mean reaction
time (or its inverse, the reaction rate) are not representative for individual
reaction events, especially in the low concentration limit.

For the case in which the nested domains have an arbitrary shape with
sufficiently smooth impermeable boundaries we presented a formally exact
spectral solution of the problem and on this basis developed a general,
self-consistent approximation, which has previously been used for several
particular geometries. We thus established a general theoretical framework
which includes previous geometrical settings as particular cases and in
which all steps involved are clearly identified. This framework will be
useful and instructive for the analysis of FRT statistics in other systems.

We then demonstrated how to apply the self-consistent approximation to
the case when the domains are concentric balls, such that the
inter-domain region, in which the particle diffuses, has the form of a
spherical shell. For such a geometry we presented explicit forms of
the Laplace-transformed FRT PDF evaluated within the
self-consistent approximation and showed that it agrees exceptionally
well with the numerically-evaluated exact spectral solution.

As we mentioned in the Introduction, our model is quite realistic and appears
indeed in many systems of biophysical and biochemical interest. In particular,
it corresponds to intermediate steps in either intracellular reaction pathways
or can be seen as the initial stage in cell-to-cell communication processes
\cite{alberts}, see also the recent work \cite{race}. In view of the relevance
to these important fields, several generalisations of the analysis presented
here may be important:

(i) In many applications, a target may not be unique, as we supposed here,
but there rather exists an array of such {\clr targets,} and the particle of interest
may react with either of them in order to trigger the same desired response.
For instance, receptors on the cellular membrane may be present in sufficiently
big amounts (see, e.g., \cite{alberts}) and binding to either of them will
result in the launch of the second messenger. The general form of the
self-consistent approximation derived here allows for an immediate extension
to this setting. Actually, the only change consists in replacing the matrix
$\K$ (which is determined by the shape of the target), and the elements of
this matrix were calculated in \cite{Grebenkov20b} for the case of multiple
non-overlapping targets of circular shape.

(ii) In some instances, the target may themselves perform a slow
diffusive motion along the boundary of the domain. Recently, some analytical
approaches have been developed to study the survival of a diffusive particle
in the presence of diffusive mobile sites in unbounded domains \cite{bray,osh,
pascal,satya,osha,LeVot}, which ultimately gives an access to the first-reaction
time PDF. An extension of these approaches to the case in which a particle
diffuses within the inter-boundary region while the target perform
diffusive motion along the boundaries represents an interesting problem in its
own right.

(iii) At some stages of the intercellular signalling, the signal can be
"amplified" \cite{alberts}. In other words, instead of a single particle,
initially some amount $N$ of identical particles are launched. This is an
important aspect when an extreme event in which the arrival of the first out
of $N$ particles matters. The impact of such an extreme event on the
FRTs and their PDF has been rather extensively discussed within the recent
years, and several analytical analyses have been proposed (see, e.g.,
\cite{N1,N2,N3,N4,N5} and references therein). An extension of such
analyses for the geometrical setting studied here should be of interest.

(iv) Finally we mention that inhomogeneities of the environment should
ultimately be included in the description. Such conditions may either
be represented by position-dependent diffusivities, by local patches
with their diffusion coefficient, or lead to anomalous diffusion
statistics. We also mention scenarios of two-channel diffusion of a
particle, that switches between inert and reactive "channels" with two
different probabilities \cite{aljaz1}, or more general switching
diffusion models that may account for reversible binding to other
molecules or organelles \cite{Grebenkov19d,Lanoiselee18,Reva21}.
{\clr Yet another extension consists in the incorporation of radial active
transport on concentric cytoskeletal tracks (e.g. microtubules), in
which the signal molecules are transported by a mixed
diffusive-convective motion (with convective motion in radial
directions).}  All of these points will deserve attention in future
work.

\ack

DG acknowledges the Alexander von Humboldt Foundation for support within
a Bessel Prize award.  RM acknowledges the German Science Foundation (DFG,
grant no. ME 1535/12-1) for support. RM also acknowledges the Foundation for
Polish Science (Fundacja na rzecz Nauki Polskiej, FNP) for support within
an Alexander von Humboldt Honorary Polish Research Scholarship.

\appendix

\section{Validation of the self-consistent approximation}
\label{sec:comparison}

In this Appendix, we discuss the accuracy of the self-consistent
approximation {\clr in Laplace domain.}

\begin{figure}
\centering
\includegraphics[width=70mm]{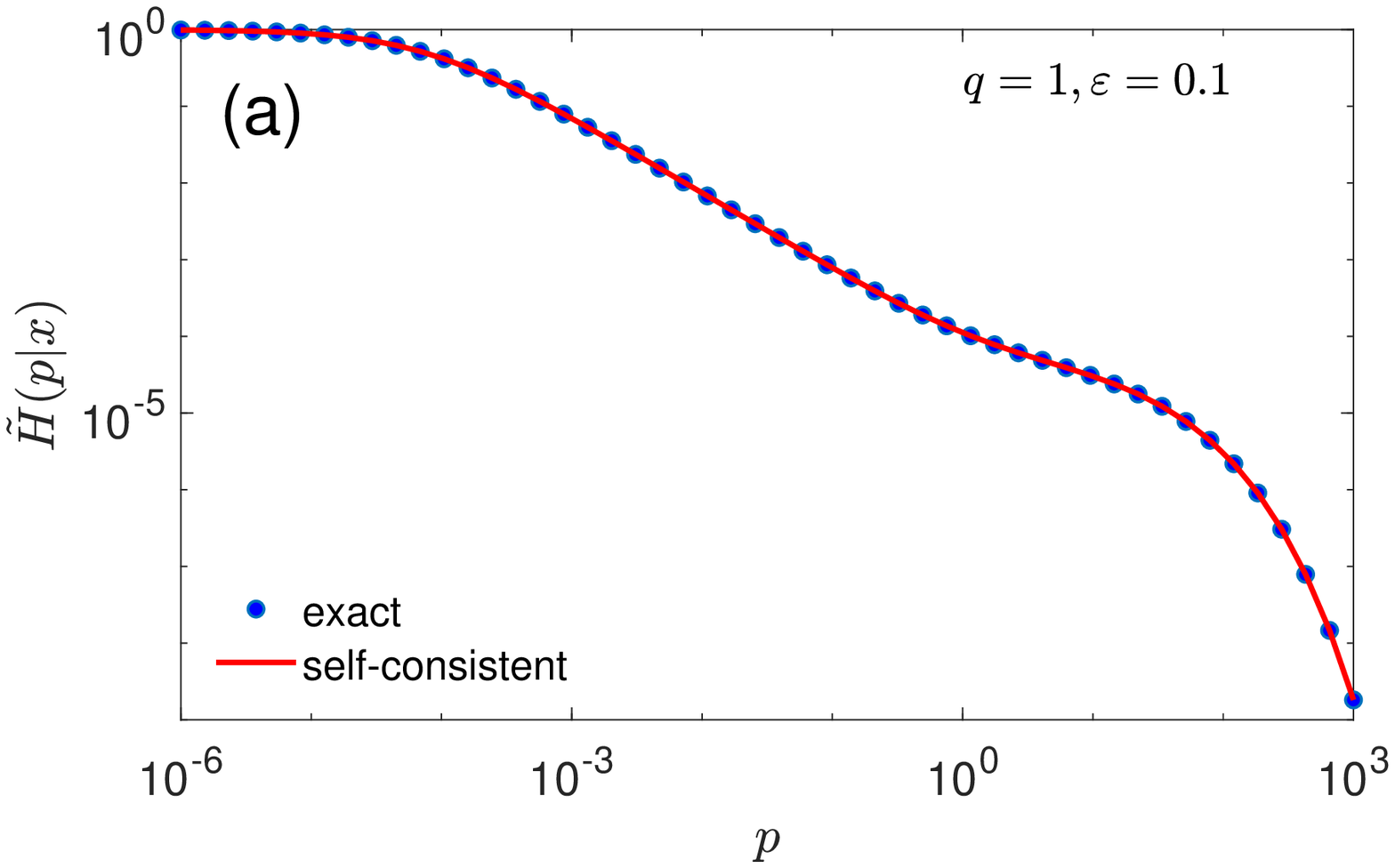}
\includegraphics[width=70mm]{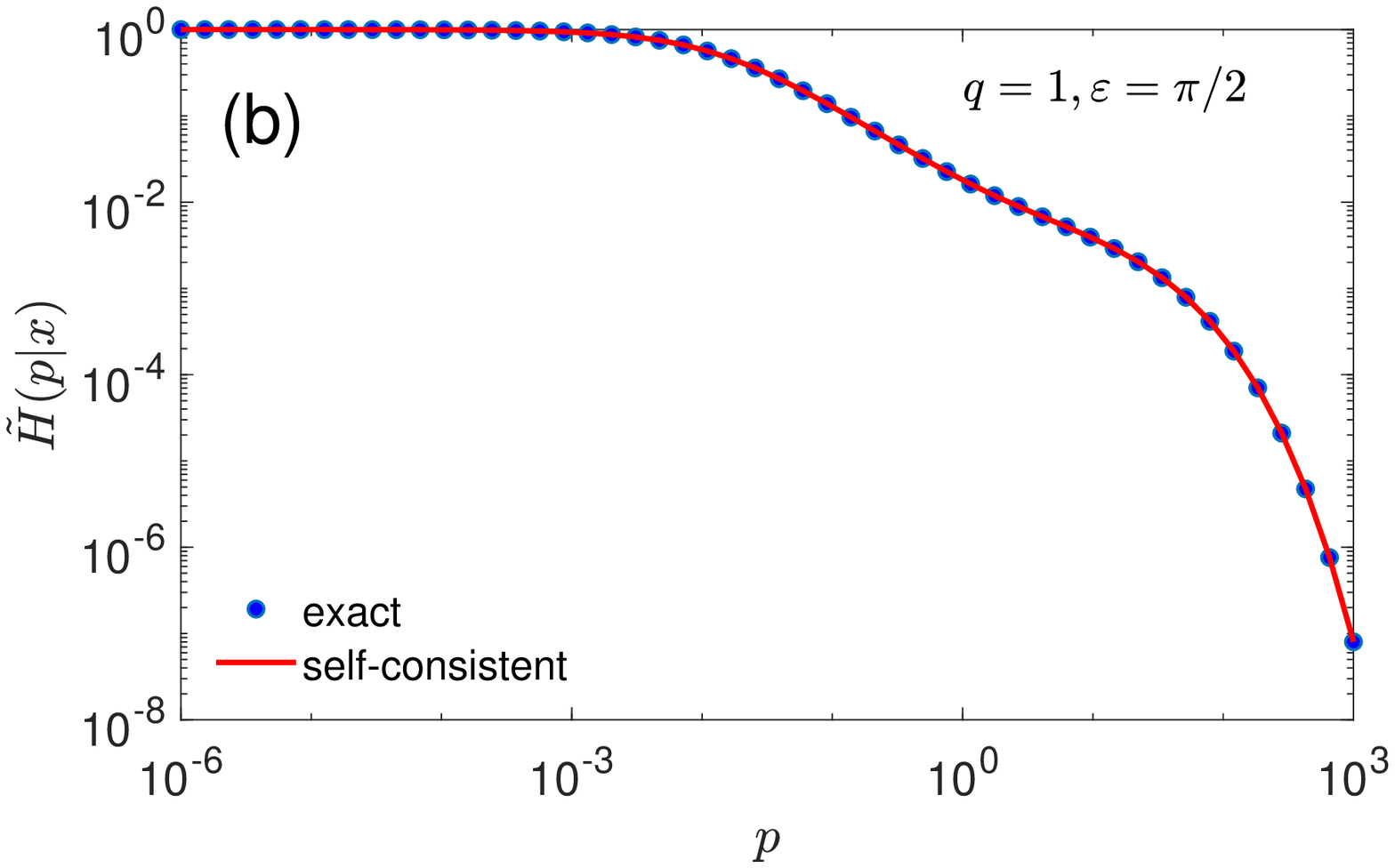}
\includegraphics[width=70mm]{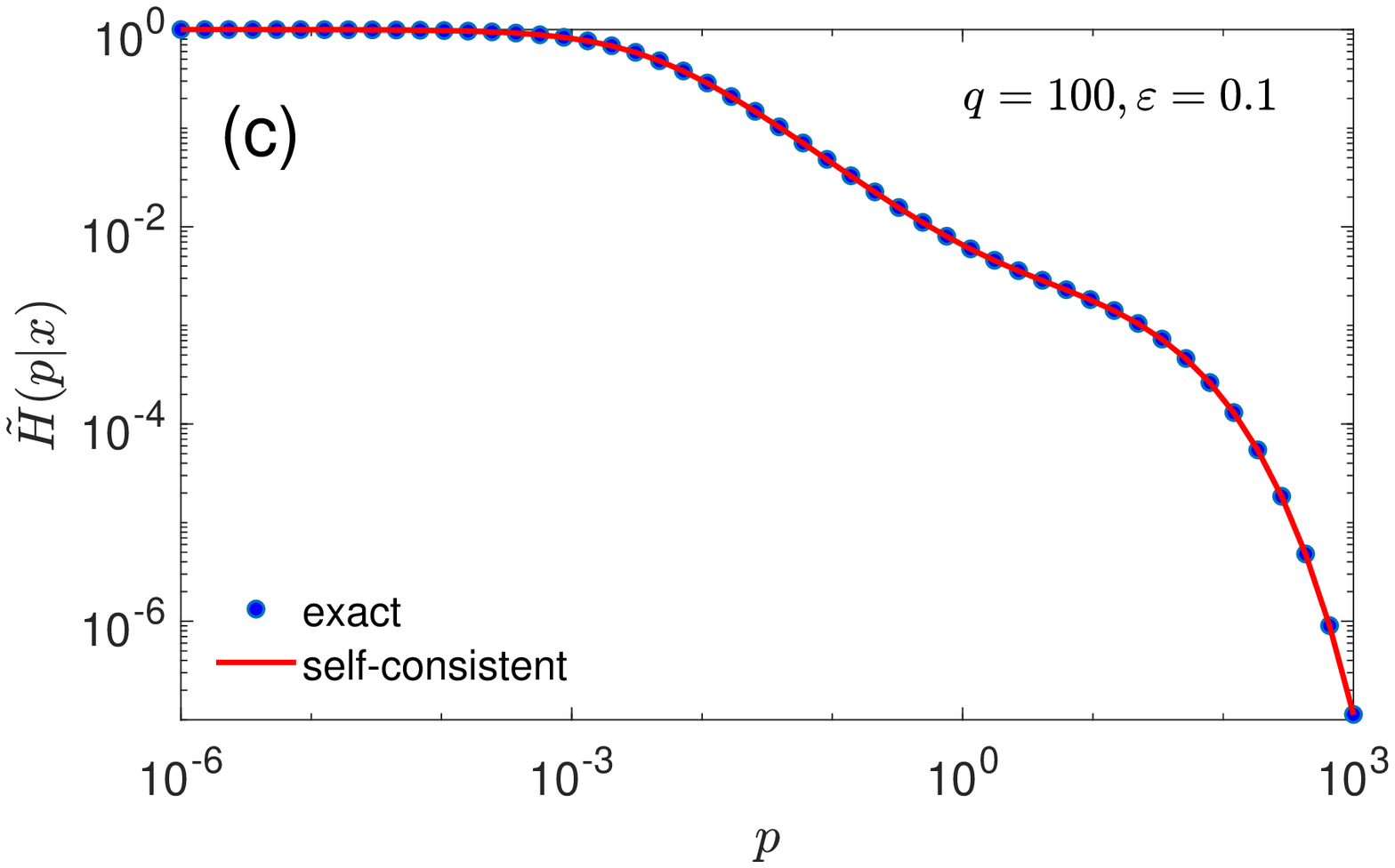}
\includegraphics[width=70mm]{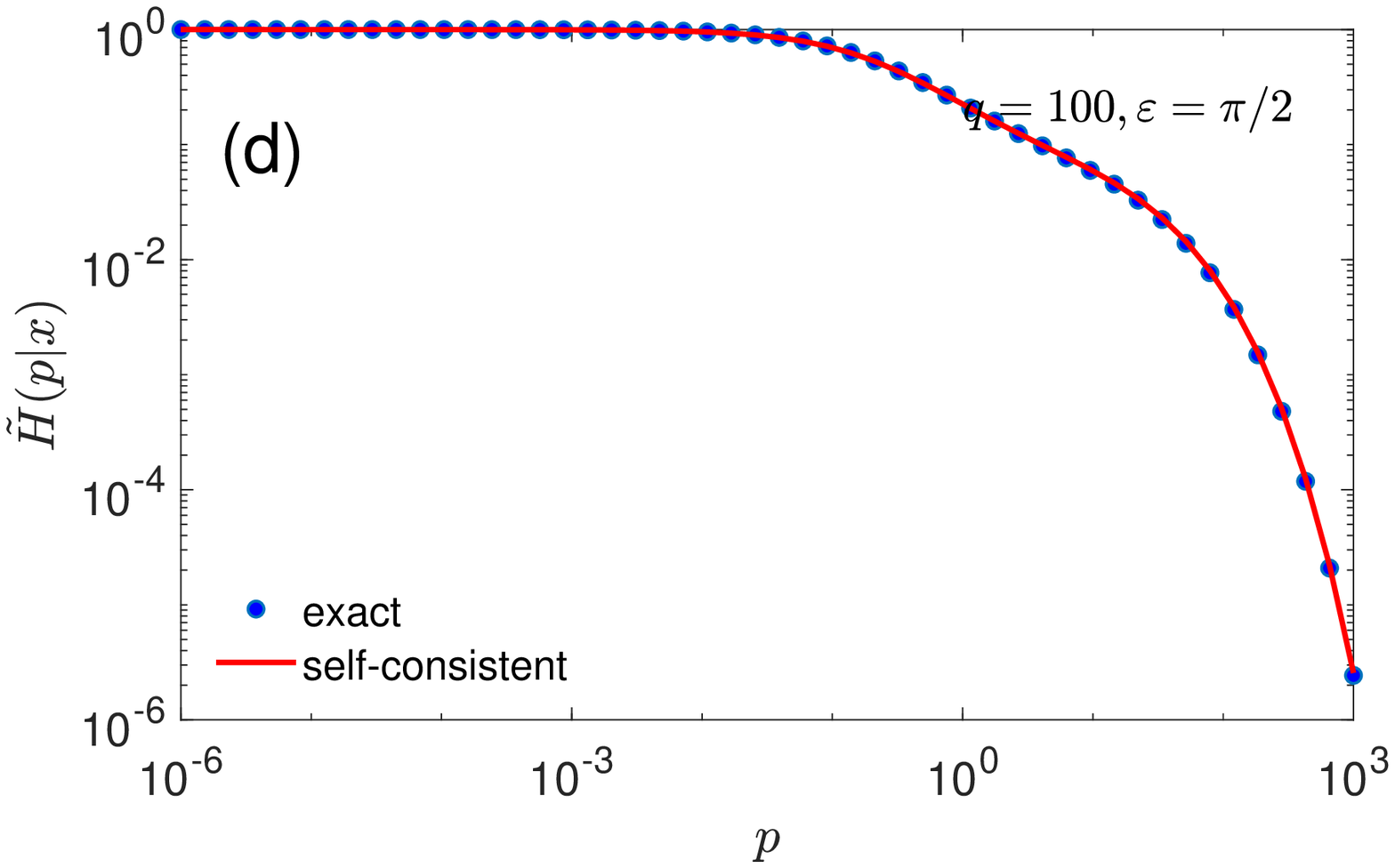}
\caption{Comparison between the SCA result \eqref{eq:Happ_inner} truncated at
$n_{\rm max}=25$ (solid line) and the exact solution
\eqref{eq:Hp_spectral2} shown by filled circles, in which the
matrices $\Mu^{(p)}$ and $\K$ are truncated at the size $(n_{\rm
max}+1)\times(n_{\rm max}+1)$. The target is located on the inner
sphere. The parameters are: $R_1=0.1$, $R_2=1$, $r=0.45$, $\theta
= 0$, and $D=1$.  {\bf (a)} $q=1$, $\ve=0.1$; {\bf (b)} $q=1$,
$\ve=\pi/2$; {\bf (c)} $q=100$, $\ve=0.1$; {\bf (d)} $q=100$, $\ve
=\pi/2$.}
\label{fig:Hp_inner}
\end{figure}

\begin{figure}
\centering
\includegraphics[width=70mm]{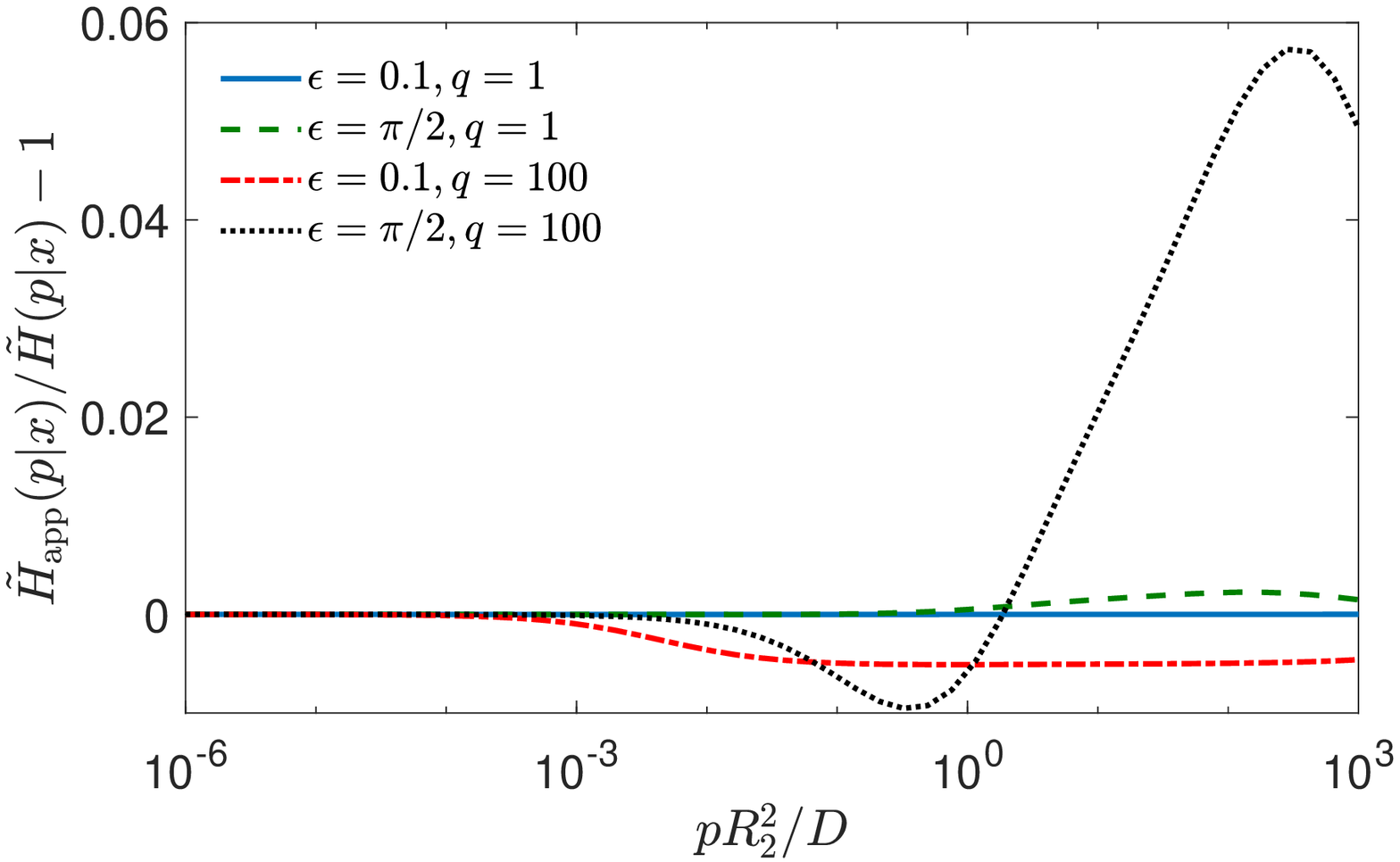}
\caption{Relative error of the SCA result \eqref{eq:Happ_inner} as compared
to the exact spectral solution. Parameters are the same as shown in figure
\ref{fig:Hp_inner}.}
\label{fig:Hp_inner_error}
\end{figure}

We start by considering the case when the target is located on the
inner sphere.  Figure \ref{fig:Hp_inner} illustrates the remarkable
agreement between the prediction \eqref{eq:Happ_inner} of the SCA and
the exact solution \eqref{eq:Hp_spectral2}. Note that both expressions
were truncated at $n_{\rm max}=25$, i.e., the size of the
(infinite-dimensional) matrices $\Mu^{(p)}$ and $\K$ was set to
$(n_{\rm max}+1)\times (n_{\rm max}+1)$, and also in
\eqref{eq:Happ_inner} we truncated the series at $n=25$. We verified
that this truncation does not affect the accuracy of the exact
solution (e.g., an increase of $n_{\rm max}$ from $25$ to $50$ does
not change the solution visibly). The two left panels {\bf (a)} and
{\bf (c)} show $\tilde{H}(p|\x)$ for a small target ($\ve=0.1$) with
moderate ($q=1$) and high ($q =100$) reactivity,
respectively. Expectedly the SCA appears to be very accurate for small
targets, regardless of the actual value of $q$. Most surprisingly,
however, the SCA turns out to be accurate even for a large target
($\ve=\pi/2$, i.e., half of the inner sphere), see the right panels
{\bf (b)} and {\bf (d)}. We note that the starting point is
relatively far from the target which is the most "favourable"
configuration for the applicability of the SCA. Yet, this is also a
typical situation in many applications. For instance, for the signal
transduction processes a particle starts on one of the boundaries and
has to reach a target on the other boundary, i.e., it appears
initially quite far from the target. The relative error of the
self-consistent approximation is shown in figure
\ref{fig:Hp_inner_error} for the same set of parameters. Expectedly,
the error is smaller when the target is smaller and less reactive. But
even in the worst case of a large, highly reactive target
($\ve=\pi/2$, $q=100$), the relative error remains below $5\%$ for a
very broad range of $p$-values, in spite of the fact that
$\tilde{H}(p|
\x)$ varies over more than $10$ orders of magnitude.

\begin{figure}
\begin{center}
\includegraphics[width=70mm]{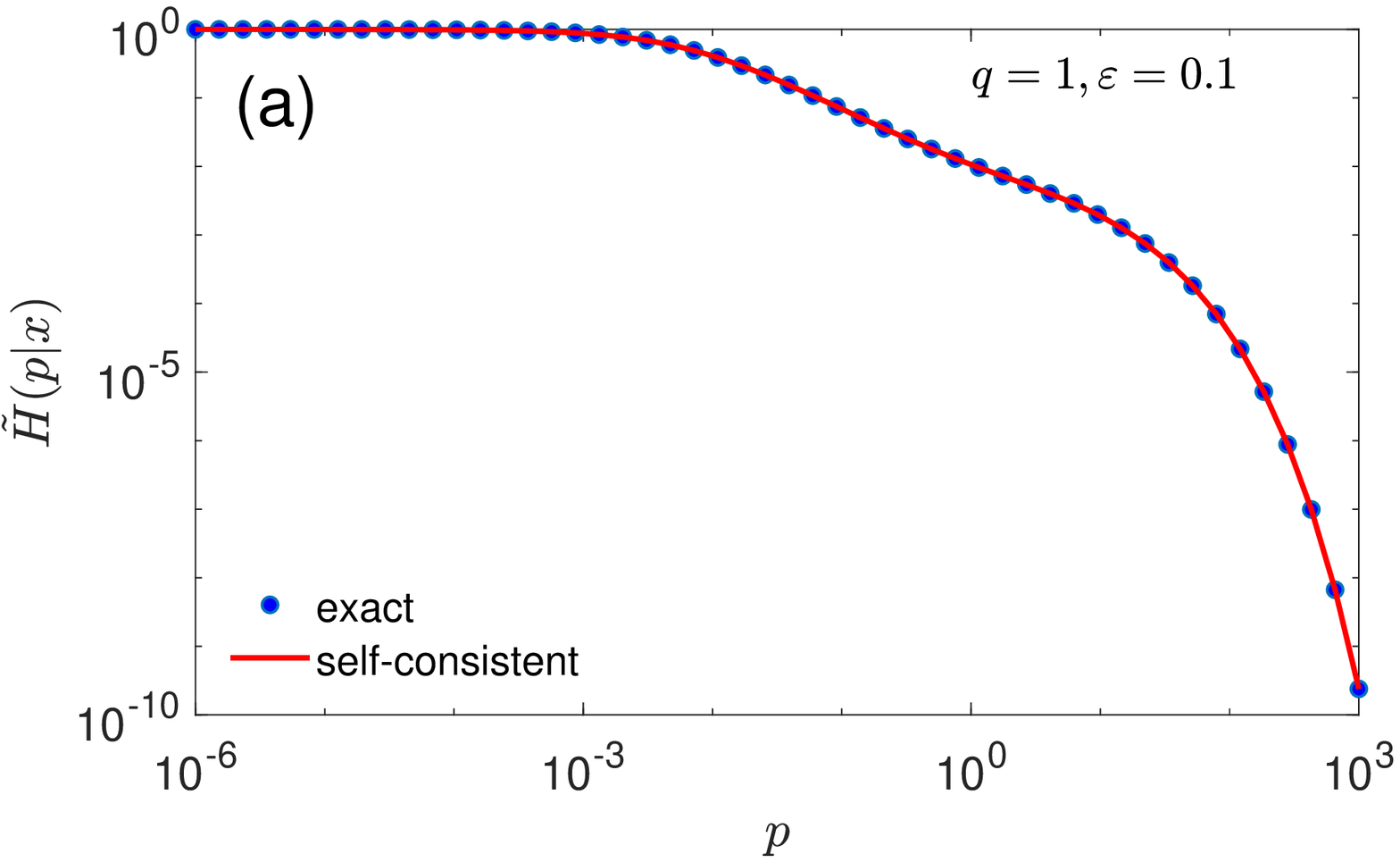}
\includegraphics[width=70mm]{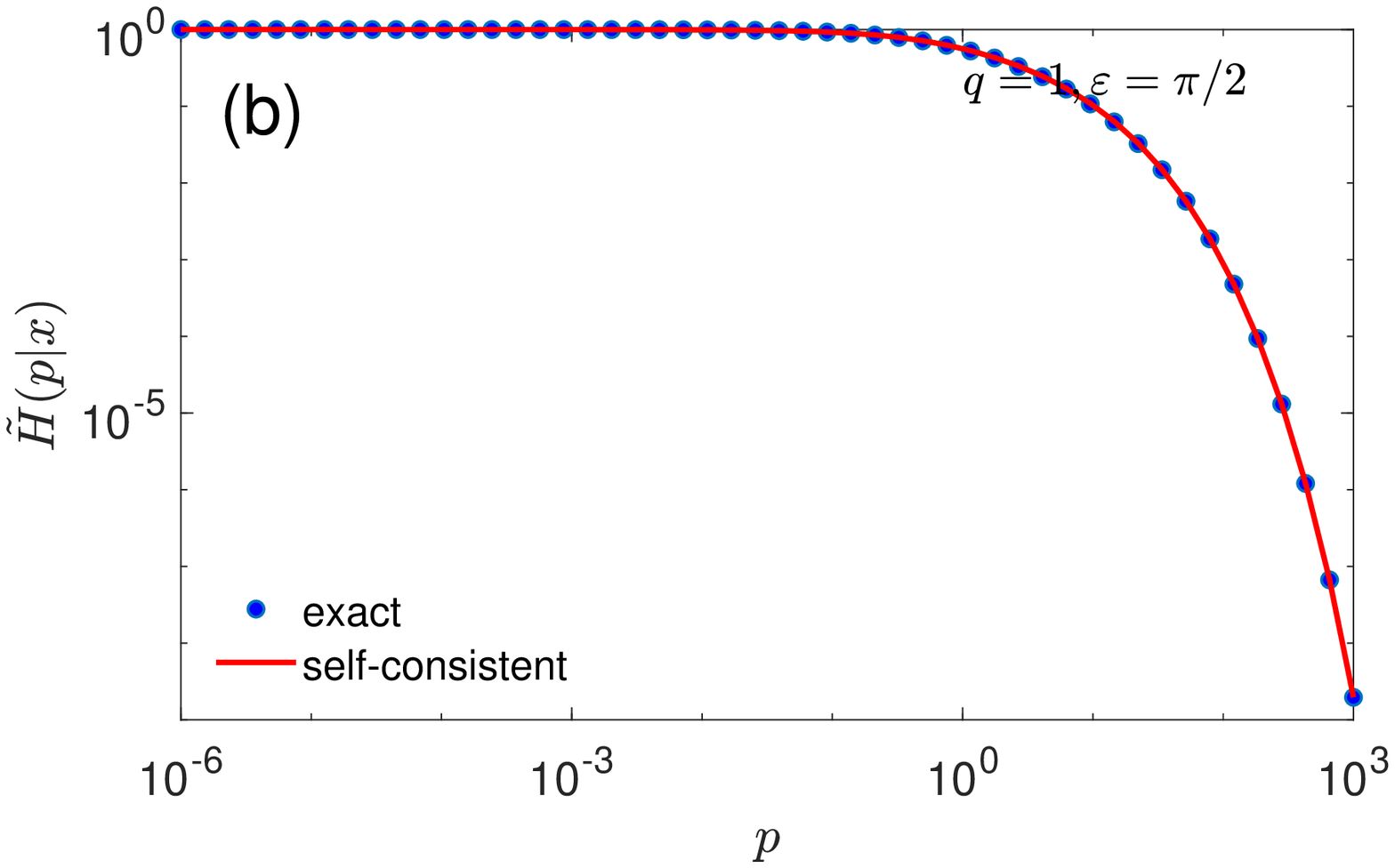}
\includegraphics[width=70mm]{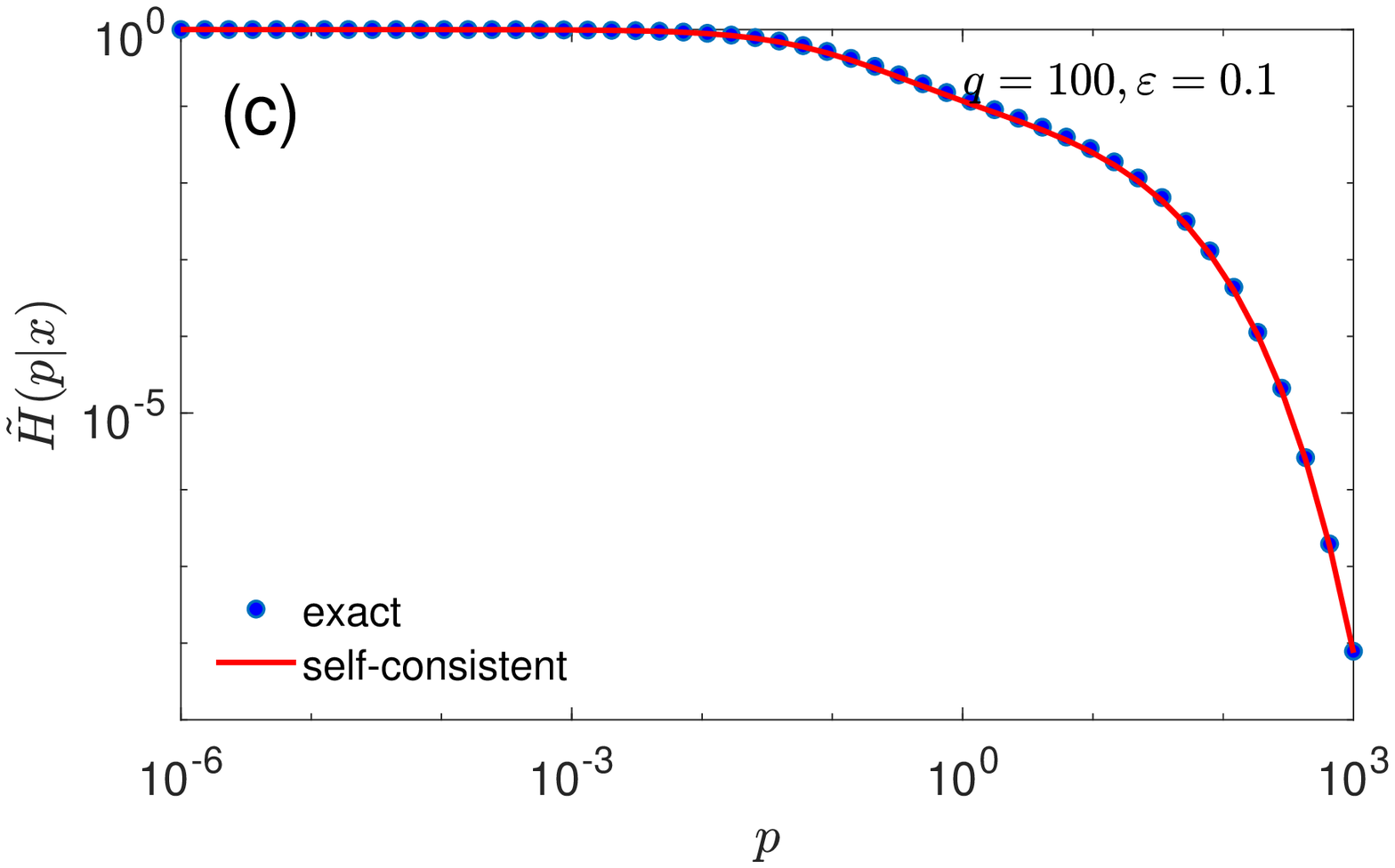}
\includegraphics[width=70mm]{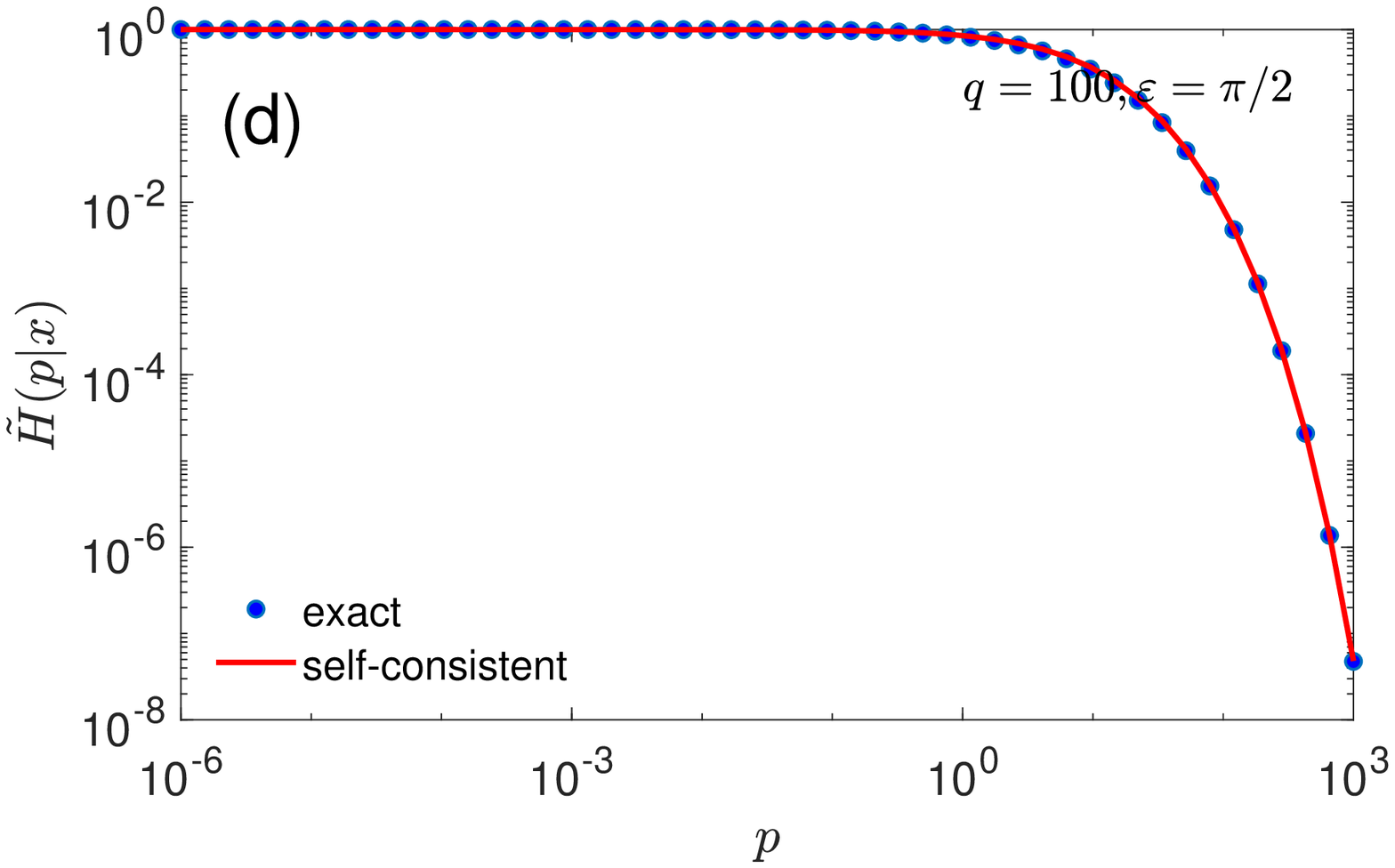}
\end{center}
\caption{Comparison between the SCA result \eqref{eq:Happ_inner} truncated at
$n_{\rm max}=25$ (solid line) and the exact solution
\eqref{eq:Hp_spectral2} shown by filled circles, in which the matrices
$\Mu^{(p)}$ and $\K$ are truncated at the size $(n_{\rm
max}+1)\times(n_{\rm max}+1)$. The target is located at the outer
sphere.  The parameters are: $R_1=0.1$, $R_2=1$, $r=0.45$, $\theta = 0$,
and $D=1$. {\bf (a)} $q=1$, $\ve=0.1$; {\bf (b)} $q=1$, $\ve=\pi/2$;
{\bf (c)} $q=100$, $\ve=0.1$; {\bf (d)} $q=100$, $\ve=\pi/2$.}
\label{fig:Hp_outer}
\end{figure}

\begin{figure}
\begin{center}
\includegraphics[width=70mm]{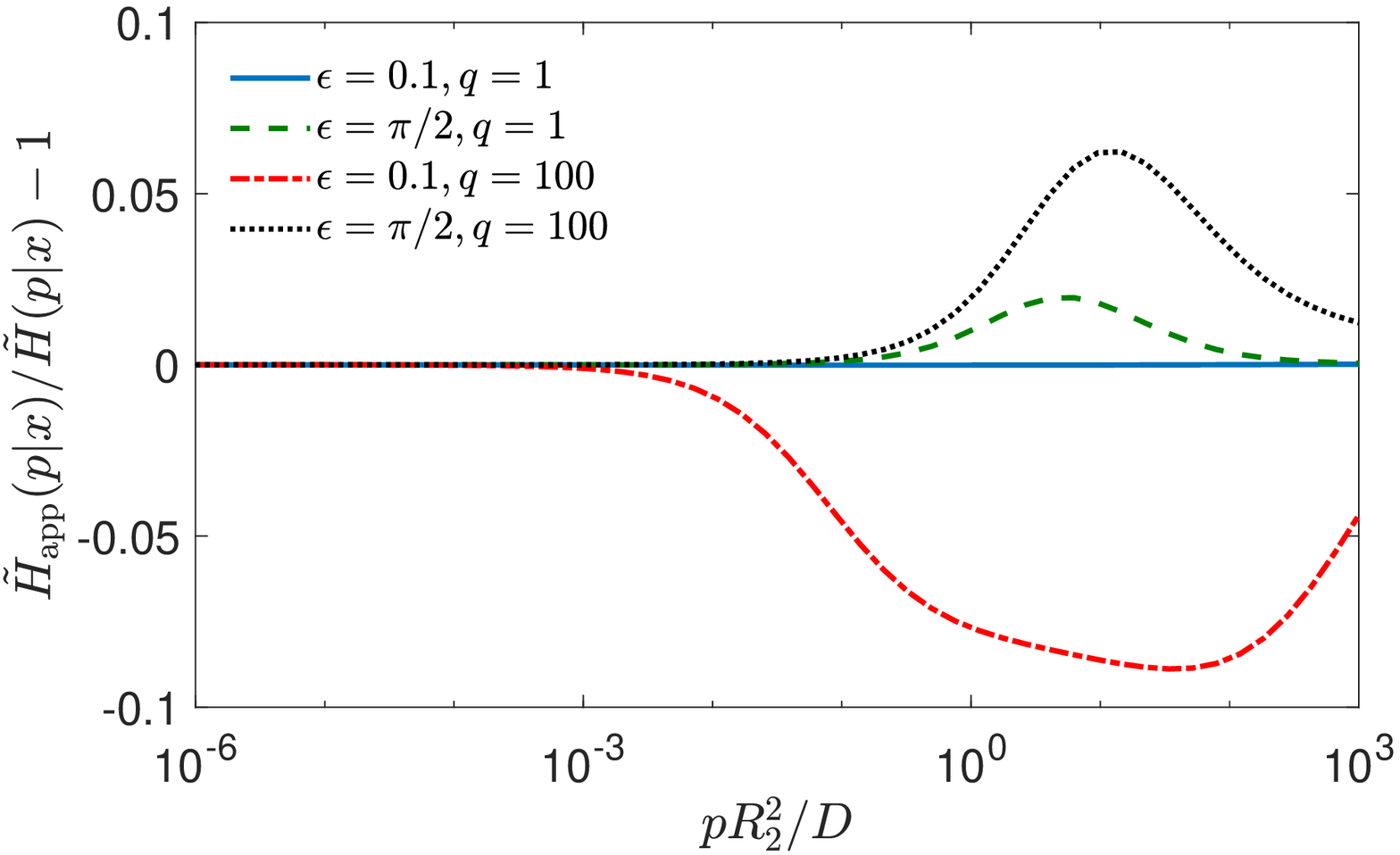}
\end{center}
\caption{Relative error of the SCA result \eqref{eq:Happ_inner} as compared to
the exact spectral solution. Parameters are the same as shown in figure
\ref{fig:Hp_outer}.}
\label{fig:Hp_outer_error}
\end{figure}

The accuracy of the SCA remains excellent for Problem II when the
target is located on the outer boundary.  Figure
\ref{fig:Hp_outer} compares the SCA prediction against the exact
spectral solution, whereas figure \ref{fig:Hp_outer_error} illustrates
its relative error.  The agreement is again remarkably good, with the
relative error raising only up to $8\%$ for large values of $p$, which
corresponds to the short-$t$ behaviour of the PDF.

\section{Mean first-reaction time}
\label{sec:MFPT}

The moments of the first-reaction time can be accessed directly from the
generating function
\begin{equation}
\E_{\x}\{\tau^k\}=(-1)^k\lim\limits_{p\to0}\frac{\partial^k}{\partial p^k}
\tilde{H}(p|\x).
\end{equation}
Once again, it is more convenient to use the SCA result $\tilde{H}_{\rm app}
(p|\x)$ for this computation, i.e.,
\begin{equation}
\E_{\x}\{\tau^k\}\approx T_k(\x)=(-1)^k\lim\limits_{p\to0}\frac{\partial^k}{
\partial p^k}\tilde{H}_{\rm app}(p|\x).
\end{equation}
Since the domain is bounded, the smallest eigenvalue $\mu_0^{(p)}$ vanishes as
$p\to 0$ \cite{Grebenkov19c}, whereas other eigenvalues $\mu_n^{(p)}$ converge
to strictly positive limits $\mu_n^{(0)}$.  

For the target on the inner sphere, one has the following expansion
\begin{equation}
\mu_0^{(p)}=ap-bp^2+O(p^3), \quad a=\frac{R_2^3-R_1^3}{3DR_1^2},\quad b=\frac{
5R_2^6-9R_2^5 R_1+5R_1^3R_2^3-R_1^6}{45D^2R_1^3},
\end{equation}
while \cite{Grebenkov20c}
\begin{equation}  \label{eq:mun0_inner}
\mu_n^{(0)}=\frac{n(n+1)}{R_1}\frac{1-(R_1/R_2)^{2n+1}}{n+(n+1)(R_1/R_2)^{2n+1}}
\quad (n>0).
\end{equation}
As a consequence, we find
\begin{equation}
J=J_1p-J_2p^2+O(p^3),\qquad J_1=\frac{a}{d},\quad J_2=\frac{a^2c+bd}{d^2},
\end{equation}
where $d=(1-\cos\ve)/2$ and 
\begin{equation}
c=\frac{1}{q}+\frac{1}{2(1-\cos\ve)}\sum\limits_{n=1}^\infty\frac{(P_{n-1}(
\cos\ve)-P_{n+1}(\cos\ve))^2}{(2n+1)\mu_n^{(0)}}.
\end{equation}
Using
\begin{equation}
\frac{g_0(r)}{\mu_0^{(p)}}=\frac{1}{ap}+e+O(p),
\end{equation}
with
\begin{equation}
e=R_1^2\frac{5r^3+10R_2^3-3r(R_1^2+6R_2^2)}{10r(R_2^3-R_1^3)}+\frac{9R_1^4R_2^2
(R_2-R_1)}{5(R_2^3-R_1^3)^2},
\end{equation}
we finally get
\begin{equation}
\label{eq:Tmean}
T_1(\x)=\frac{ac}{d}+\frac{b}{a}-ae-\frac{a}{d}\sum\limits_{n>0}\frac{g_n^{(p=0)}
(r)}{\mu_n^{(0)}}P_n(\cos\theta)\frac{P_{n-1}(\cos\ve)-P_{n+1}(\cos\ve)}{2},
\end{equation}
where
\begin{equation}
g_n^{(p=0)}(r)=(R_1/r)^{n+1}\frac{n+(n+1)(r/R_2)^{2n+1}}{n+(n+1)(R_1/R_2)^{2n+1}}.
\end{equation}

For the target on the outer sphere, one has
\begin{equation}
\mu_0^{(p)}=ap-bp^2+O(p^3),\quad a=\frac{R_2^3-R_1^3}{3DR_2^2},\quad b=\frac{R_2^6
-5R_1^3R_2^3+9R_2 R_1^5-5R_1^6}{45D^2R_2^3},
\end{equation}
while \cite{Grebenkov20c}
\begin{equation}
\mu_n^{(0)}=\frac{n(n+1)}{R_2}\frac{1-(R_1/R_2)^{2n+1}}{n+1+n(R_1/R_2)^{2n+1}}
\quad(n>0).
\end{equation}
As a consequence, we find that
\begin{equation}
J=J_1p-J_2p^2+O(p^3),\qquad J_1=\frac{a}{d},\quad J_2=\frac{a^2c+bd}{d^2},
\end{equation}
with $c$ and $d$ as defined above. Using
\begin{equation}
\frac{g_0(r)}{\mu_0^{(p)}}=\frac{1}{ap}+e+O(p),
\end{equation}
with
\begin{equation}
e=R_2^2\frac{5r^3+10R_1^3-3r(R_2^2+6R_1^2)}{10r(R_2^3-R_1^3)}+\frac{9R_1^2R_2^4
(R_2-R_1)}{5(R_2^3-R_1^3)^2},
\end{equation}
we get again equation \eqref{eq:Tmean} for $T_1(\x)$, with
\begin{equation}
g_n^{(p=0)}(r)=(r/R_2)^n\frac{n+1+n(R_1/r)^{2n+1}}{n+1+n(R_1/R_2)^{2n+1}}.
\end{equation}

\section{A planar circular annulus domain}
\label{sec:planar}

The general approach from section \ref{sec:general} can also be applied to a
planar circular annulus domain $\Omega=\{\x\in\R^2 ~:~ R_1<|\x|<R_2\}$, with
the target represented by an arc $(-\ve,\ve)$ either on the inner circle, or
on the outer circle. This problem is also equivalent to diffusion between two
coaxial cylinders of radii $R_1$ and $R_2$, (i.e., in the domain $\Omega'=
\Omega\times\R\subset \R^3$), towards a vertical infinitely long partially
reactive stripe of angular size $2\ve$. Note that this shape of the target
is different from that considered in \cite{11,dist5,dist4}.

In this case, the eigenbasis of the Dirichlet-to-Neumann operator is well known.
For instance, for the target on the inner circle, one has \cite{Grebenkov20c}:
$v_n^{(p)}=e^{in\theta}/\sqrt{2\pi R_1}$ and $\mu_n^{(p)}=-g'_n(R_1)$, where
\begin{equation}
\label{eq:gnI_2D}
g_n(r)=\frac{K'_n(\alpha R_2)I_n(\alpha r)-I'_n(\alpha R_2)K_n(\alpha r)}{K'_n(
\alpha R_2)I_n(\alpha R_1)-I'_n(\alpha R_2)K_n(\alpha R_1)},
\end{equation}
with $n\in \Z$. One then easily gets
\begin{equation}
\K_{n,n'}=\int\limits_{-\ve}^{\ve}d(R_1\theta)\frac{e^{i(n-n')\theta}}{2\pi R_1}
=\frac{\sin(n-n')\ve}{\pi(n-n')}  
\end{equation}
and
\begin{equation}
C_n=\int\limits_{-\ve}^{\ve}d(R_1\theta)\frac{e^{in\theta}}{\sqrt{2\pi R_1}}=
\sqrt{2\pi R_1}\frac{\sin n\ve}{\pi n}.
\end{equation}
The exact solution \eqref{eq:Hp_spectral} is then
\begin{equation}
\label{eq:Hp_spectral_2D}
\tilde{H}(p|\x)=\sum\limits_{n,n'=-\infty}^\infty e^{in\theta}\biggl[\bigl(\Mu
^{(p)}/q+\K^{(p)}\bigr)^{-1}\biggr]_{n,n'}\frac{\sin n'\ve}{\pi n'}\qquad(\x\in\pa),
\end{equation}
while the self-consistent approximation reads
\begin{equation}
\label{eq:Happ_2D}
\tilde{H}_{\rm app}(p|\x)=J\sum\limits_{n=-\infty}^\infty g_n(r)e^{in\theta}
\frac{1}{\mu_n^{(p)}}\frac{\sin n\ve}{\pi n},
\end{equation}
with
\begin{equation}
\label{eq:J_2D}
J=\left(\frac{1}{q}+\frac{\pi}{\ve}\sum\limits_{n=-\infty}^\infty\frac{1}{\mu_n^{
(p)}}\biggl(\frac{\sin n\ve}{\pi n}\biggr)^2\right)^{-1} .
\end{equation}
Note that the above summations can be reduced to nonnegative values of $n$ by
symmetry.

When the target is on the outer circle one has $v_n^{(p)}=e^{in\theta}/\sqrt{2
\pi R_2}$ and $\mu_n^{(p)}=g'_n(R_2)$, and 
\begin{equation}
\label{eq:gnII_2D}
g_n(r)=\frac{K'_n(\alpha R_1)I_n(\alpha r)-I'_n(\alpha R_1)K_n(\alpha r)}{K'_n(
\alpha R_1)I_n(\alpha R_2)-I'_n(\alpha R_1)K_n(\alpha R_2)},
\end{equation}
while relations \eqref{eq:Hp_spectral_2D}, \eqref{eq:Happ_2D}, and \eqref{eq:J_2D}
remain unchanged.

\section{\clr Spread harmonic measure}
\label{sec:spread}

\begin{figure}
\begin{center}
\includegraphics[width=70mm]{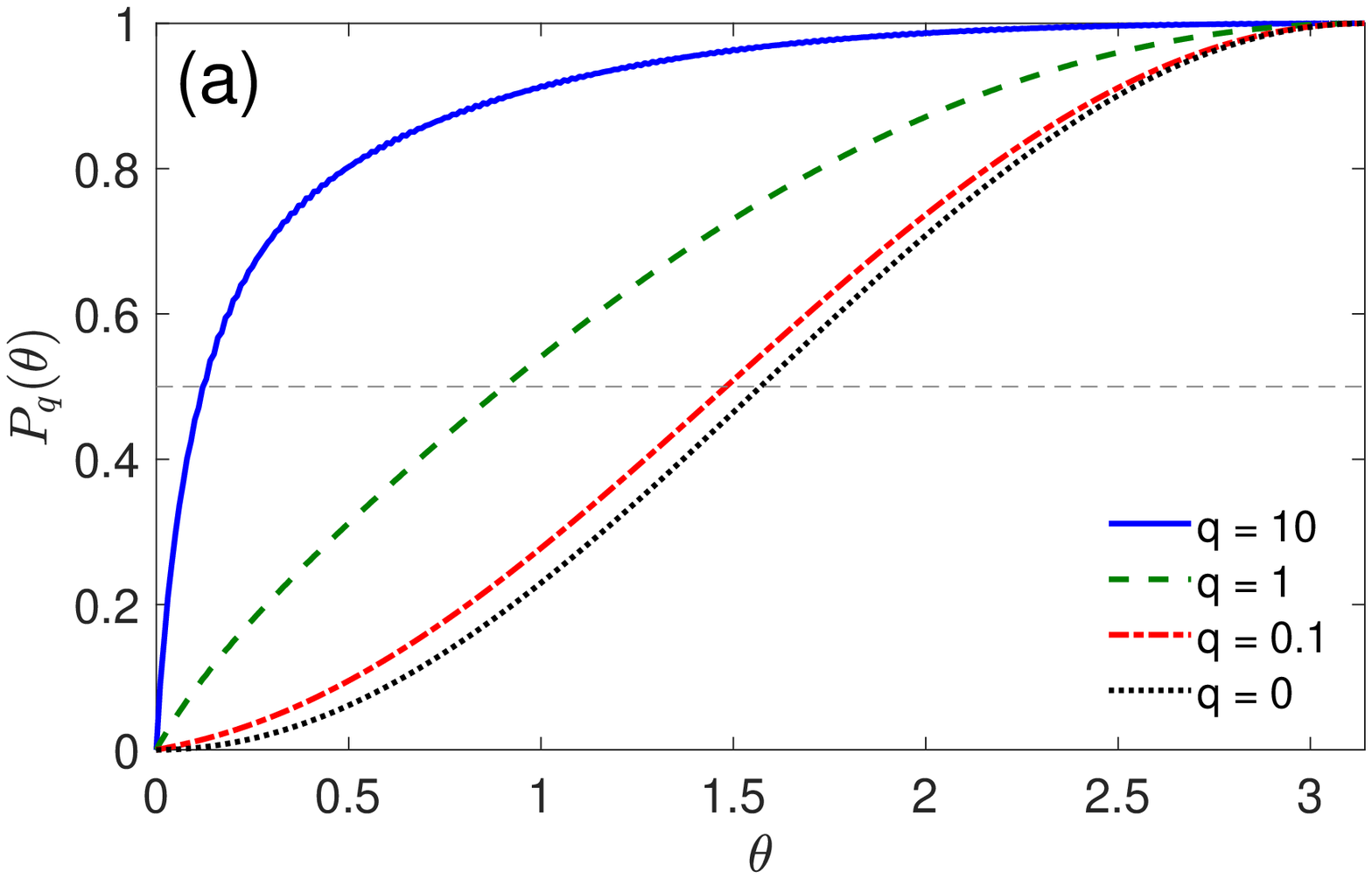}
\includegraphics[width=70mm]{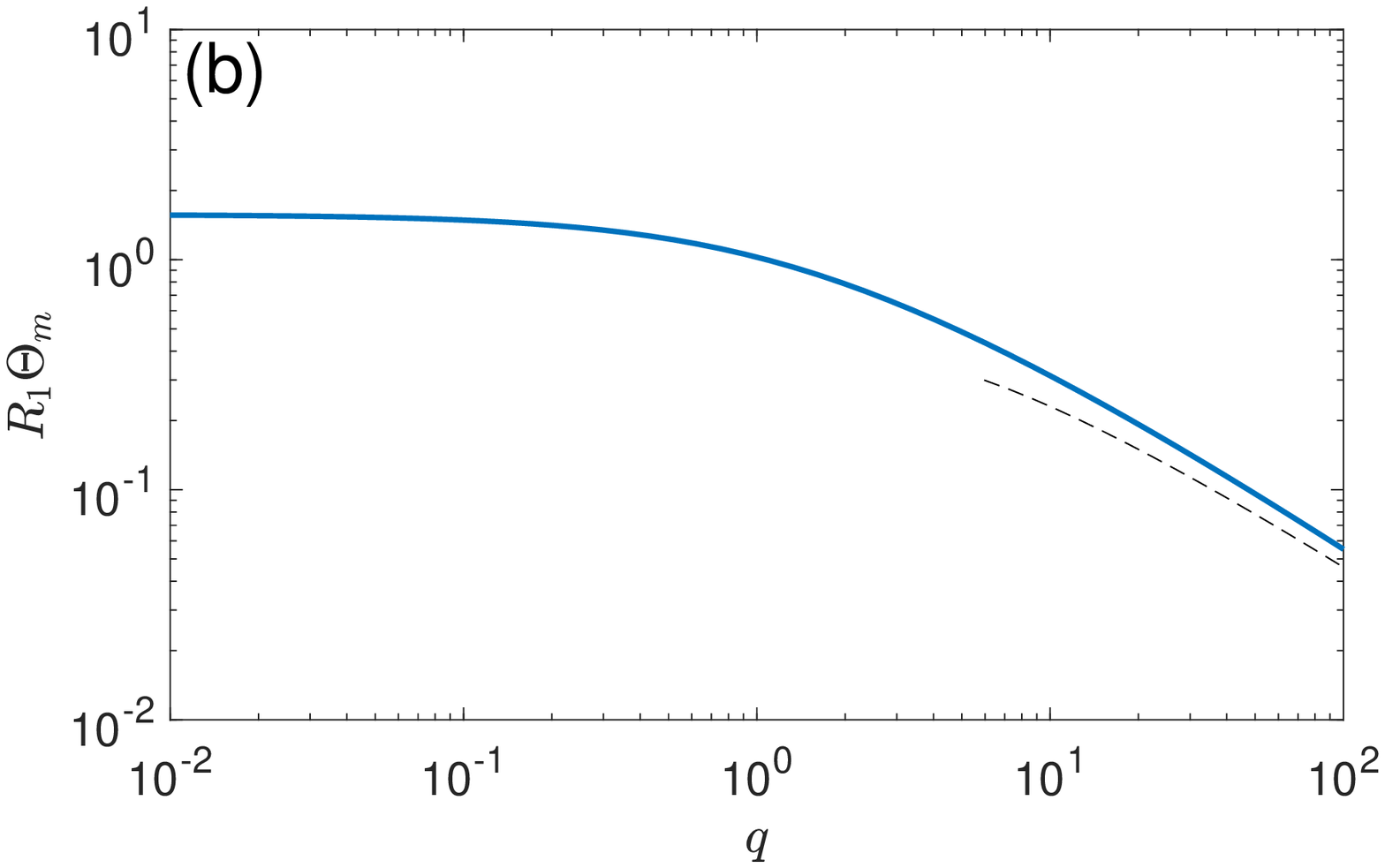}
\end{center}
\caption{\clr {\bf (a)} Probability $P_q(\theta)$ for the reaction event to occur
on the inner sphere within the spherical cap of angle $\theta$, for four values
of $q$, with $R_1=1$ and $R_2=2$. Note that $P_0(\theta)=(1-\cos\theta)/2$.
The crossing with the horizontal line at $1/2$ determines the size $R_1\Theta_t$
of the typical region on which the half of reaction events occurs. For large $q$,
$R_1\Theta_t\propto1/q$. {\bf (b)} Mean geodesic distance $R_1\Theta_m$ as
function $q$, for $R_1=1$ and $R_2=2$. The dashed line shows $q\ln(q)$.}
\label{fig:Theta}
\end{figure}

{\clr 

In this Appendix, we provide some insights into the role of the parameter $q$
characterising the reactivity of the target. For the sake of brevity, we focus
on the case when the target covers the whole inner sphere.  After the first
arrival onto the inner sphere, the particle does not necessarily react
immediately but may be reflected to continue its bulk diffusion, return to the
target and be reflected again, and so on, until an eventual reaction after a
number of failed attempts. As a consequence, the location of the reaction event
is in general different from the location of the first arrival. The distribution
of reaction locations is called the spread harmonic measure \cite{Grebenkov06c}.
Some properties of this measure and the related interpretation of the parameter
$q$ were discussed in \cite{Sapoval05,Grebenkov06c,Grebenkov15c} for a restricted
class of domains. In general, the spread harmonic measure density $\omega_q(\s|
\s_0)$, characterising the probability of the reaction event at a boundary point
$\s$ after the arrival onto the boundary point $\s_0$, can be expressed in terms
of the eigenfunctions of the Dirichlet-to-Neumann operator \cite{Grebenkov20}
\begin{equation}
\omega_q(\s|\s_0)=\sum\limits_{n=0}^\infty\frac{v_n^{(0)}(\s)[v_n^{(0)}(\s_0)]^*
}{1 + \mu_n^{(0)}/q}.
\end{equation}
For the considered shell-like domain in which the target covers the whole inner
sphere, the first arrival point $\s_0$ can be set to be on the North pole,
without loss of generality. The explicit form \eqref{eq:vn_inner} of the
eigenfunctions yields
\begin{equation}
\omega_q(\s|\s_0)=\frac{1}{4\pi R_1^2}\sum\limits_{n=0}^\infty\frac{(2n+1)P_n(
\cos\theta)}{1+\mu_n^{(0)}/q},
\end{equation}
with $\mu_n^{(0)}$ given by \eqref{eq:mun0_inner} for $n > 0$, and $\mu_0^{(0)}
=0$. Integrating over the azimuthal angle $\phi$ and multiplying by $R_1^2$, we
get the distribution of the polar angle $\theta$ of the reaction point,
\begin{equation}
\omega_q(\theta)=\frac{\sin\theta}{2}\sum\limits_{n=0}^\infty\frac{(2n+1)P_n(
\cos\theta)}{1+\mu_n^{(0)}/q},
\end{equation}
where $\sin\theta$ comes from using spherical coordinates. The integral of
$\omega_q(\theta')$ from $0$ to $\theta$ is the probability that the reaction
location is on the spherical cap of angle $\theta$,
\begin{equation}
P_q(\theta)=\int\limits_0^{\theta}d\theta'\omega_q(\theta')=\frac{1}{2} 
\sum\limits_{n=0}^\infty\frac{P_{n-1}(\cos\theta)-P_{n+1}(\cos\theta)}{1+
\mu_n^{(0)}/q}.
\end{equation}
Setting $P_q(\Theta_t)=1/2$ determines implicitly the angular size $\Theta_t$
of the spherical cap, on which half of the reaction events occur. Figure
\ref{fig:Theta}(a) illustrates the behaviour of this probability for three
values of $q$.

In addition, one can calculate the mean angle $\Theta_m$, which is related to
the mean geodesic distance between the North pole (the first arrival) and the
reaction location,
\begin{equation} 
\Theta_m=\int\limits_0^\pi d\theta\theta\omega_q(\theta)=\frac12\sum\limits_{
n=0}^\infty\frac{(2n+1)}{1+\mu_n^{(0)}/q}I_n,
\end{equation}
where
\begin{equation}
I_n=\int\limits_{-1}^1dx\arccos(x)P_n(x)=\pi\times\left\{\begin{array}{ll}1&
(n=0),\\
-\left(\frac{(n-2)!!}{2^{(n+1)/2}((n+1)/2)!}\right)^2&(n\textrm{ odd}),\\
0&(n\textrm{ even}).\\\end{array}\right.
\end{equation}
We thus obtain
\begin{equation}
\Theta_m=\frac{\pi}{2}\left(1-\sum\limits_{k=0}^{\infty}\frac{2(2k+1)+1}{1+\mu
_{2k+1}^{(0)}/q}c_k\right),
\end{equation}
with
\begin{equation}
c_k=\biggl(\frac{(2k-1)!!}{2^{k+1}(k+1)!}\biggr)^2\qquad(k=0,1,2,\ldots).
\end{equation}

In the limit of high reactivity, $q\to \infty$, the spread harmonic measure
density converges to a Dirac distribution around the first arrival point.
Since $c_k\propto k^{-3}$, one gets the asymptotic behaviour $\Theta_m\propto
\ln(qR_1)/(qR_1)$. In other words, for large $q$, the inverse of $q$ is
proportional to the mean geodesic distance $R_1\Theta_m$, up to a logarithmic
correction. In the opposite limit of low reactivity, $q\to0$, the spread harmonic
measure density is getting uniform, and one gets $\Theta_m \to \pi/2$. This
reflects the fact that the inner sphere, as well as all geodesic distance on it,
are bounded. In other words, the mean geodesic distance cannot grow as $1/q$
forever, as it was the case for the plane \cite{Sapoval05,Grebenkov15c}. The
dependence of $\Theta_m$ on $q$ is illustrated on figure \ref{fig:Theta}(b).}

\section*{References}

\begin{thebibliography}{999}

\bibitem{berg} Berg HC and Purcell EM 1977 {\em Physics of chemoreception}
Biophys. J. {\bf 20} 193

\bibitem{lau} Lauffenburger DA and Linderman JJ, eds., 1996 {\em
Receptors. Models for binding, trafficking, and signalling}, (IRL Press,
Oxford)

\bibitem{trans1} Purves D et al, eds., 2001 {\em Neuroscience}, 2nd ed.,
(Sunderland (MA): Sinauer Associates)

\bibitem{trans2} Bradshaw RA and Dennis EA, eds., 2010 {\em Handbook of Cell
Signaling}, 2nd ed., (Amsterdam, Netherlands: Academic Press)

\bibitem{alberts} Alberts B {\em et al} 2015 {\em Molecular Biology of the
Cell}, $6$th ed., (New York: Garland)

\bibitem{100} Hafner A and Rieger H 2018 {\em Spatial Cytoskeleton Organization
Supports Targeted Intracellular Transport} Biophys. J. {\bf 114} 1420

\bibitem{100c} Mangeat M and Rieger H {\em The Narrow Escape Problem in
two-shell spherical domains};  arXiv:2104.13125v1

\bibitem{5c} Holcman D and Schuss Z 2014 {\em The narrow escape problem}
SIAM Rev. {\bf 56} 213

\bibitem{7} Cheviakov AF,  Ward MJ and Straube R 2010 {\em An asymptotic
analysis of the mean first passage time for narrow escape problems: part II:
the sphere} Multiscale Model. Simul. {\bf 88} 36

\bibitem{7b} Cheviakov AF, Reimer AS and Ward MJ 2012 {\em Mathematical
modeling and numerical computation of narrow escape problems} Phys. Rev. E
{\bf 85} 021131

\bibitem{Caginalp12} Caginalp C and Chen X 2012 {\em Analytical and Numerical
Results for an Escape Problem} Arch. Rational. Mech. Anal. {\bf 203} 329-342

\bibitem{Marshall16} Marshall JS 2016 {\em Analytical Solutions for an
Escape Problem in a Disc with an Arbitrary Distribution of Exit Holes Along
Its Boundary} J. Stat. Phys. {\bf 165} 920-952

\bibitem{Grebenkov16c} Grebenkov DS 2016 {\em Universal formula for the
mean first passage time in planar domains} Phys. Rev. Lett. {\bf 117} 260201

\bibitem{Lindsay17} Lindsay AE, Bernoff AJ, and Ward MJ 2017 {\em First
Passage Statistics for the Capture of a Brownian Particle by a Structured
Spherical Target with Multiple Surface Traps}, Multiscale Model. Simul. {\bf
15} 74-109

\bibitem{Bernoff18a} Bernoff AJ and Lindsay AE 2018 {\em Numerical
approximation of diffusive capture rates by planar and spherical surfaces
with absorbing pores} SIAM J. Appl. Math. {\bf 78} 266-290

\bibitem{6} B\'enichou O and Voituriez R 2008 {\em Narrow-escape time problem:
time needed for a particle to exit a confining domain through a small window}
Phys. Rev. Lett. {\bf 100} 168105

\bibitem{6a} B\'enichou O and Voituriez R 2014 {\em From first-passage
times of random walks in confinement to geometry-controlled kinetics}
Phys. Rep. {\bf 539} 225

\bibitem{8} Oshanin G, Tamm M and Vasilyev O 2010 {\em Narrow-escape times
for diffusion in microdomains with a particle-surface affinity: Mean-field
results} J. Chem. Phys. {\bf 132} 06B607

\bibitem{9} B\'enichou O, Grebenkov D S, Levitz P, Loverdo C and Voituriez
R 2010 {\em Optimal reaction time for surface-mediated diffusion}
Phys. Rev. Lett. {\bf 105} 150606

\bibitem{9a} B\'enichou O, Grebenkov DS, Levitz P, Loverdo C and Voituriez R
2011 {\em Mean first-passage time of surface-mediated diffusion in spherical
domains} J. Stat. Phys. {\bf 142} 657

\bibitem{9b} Rupprecht J-F, B\'enichou O, Grebenkov D S and Voituriez R 2012
{\em Kinetics of active surface-mediated diffusion in spherically symmetric
domains} J. Stat. Phys. {\bf 147} 891

\bibitem{9c} Rupprecht J-F, B\'enichou O, Grebenkov DS, and Voituriez R 2015
{\em Exit time distribution in spherically symmetric two-dimensional domains}
J. Stat. Phys. {\bf 158} 192-230

\bibitem{11} Grebenkov DS and Oshanin G 2017 {\em Diffusive escape
through a narrow opening: new insights into a classic problem}
Phys. Chem. Chem. Phys. {\bf 19} 2723

\bibitem{mat} Berezhkovskii AM, Coppey M, Sealfon SC and Shvartsman S 2008 {\em Cell-to-cell communication: Time and length scales of ligand internalization in cultures of suspended cells}
J. Chem. Phys. {\bf 128}  225102

\bibitem{mor} Hughes N, Faulkner C,  Morris RJ and Tomkins M 2021 {\em Intercellular Communication as a Series of Narrow Escape Problems} IEEE Transactions on Molecular, Biological and Multi-Scale Communications {\bf 7}  89 

\bibitem{carlos1} Mej\'{i}a-Monasterio C, Oshanin G and Schehr G 2011 {\em
First passages for a search by a swarm of independent random searchers}
J. Stat. Mech.  P06022

\bibitem{carlos2} Mattos TG, Mej\'{i}a-Monasterio C, Metzler R and  Oshanin
G 2012 {\em First passages in bounded domains: When is the mean first passage
time meaningful?} Phys. Rev. E {\bf 86} 031143

\bibitem{redner} Redner S 2001 {\em A Guide to First Passage Processes}
(Cambridge: Cambridge University Press), p. 107.

\bibitem{redner1} Metzler R, Oshanin G and Redner S, eds., 2014 {\em
First-passage phenomena and their applications}, (World Scientific Publishers:
Singapore)

\bibitem{redner2} Lindenberg K, Metzler R and Oshanin G, eds., 2019 {\em
Chemical Kinetics: Beyond the Textbook}, (World Scientific Publishers Europe:
London)

\bibitem{Grebenkov19a} Grebenkov DS 2019 {\em Spectral theory of imperfect
diffusion-controlled reactions on heterogeneous catalytic surfaces},
J. Chem. Phys. {\bf 151} 104108

\bibitem{Grebenkov19b}  Grebenkov DS  and Traytak SD 2019 {\em Semi-analytical
computation  of Laplacian Green functions in three-dimensional domains with
disconnected spherical boundaries} J. Comput. Phys. {\bf 379} 91

\bibitem{Grebenkov20b}  Grebenkov DS 2020 {\em Diffusion toward non-overlapping
partially reactive spherical traps: fresh insights onto classic problems}
J. Chem. Phys. {\bf 152} 244108

{\clr
\bibitem{Sokolowski19}	Sokolowski TR, Paijmans J, Bossen L, Miedema T, Wehrens
M, Becker NB, Kaizu K, Takahashi K, Dogterom M, and Rein ten Wolde P 2019
{\em eGFRD in all dimensions} J. Chem. Phys. {\bf 150} 054108
}

\bibitem{szabo} Shoup D, Lipari G and Szabo A 1981 {\em Diffusion-controlled
bimolecular reaction rates. The effect of rotational diffusion and orientation
constraints} Biophys. J. {\bf 36} 697

\bibitem{colloid} Oshanin G, Popescu MN and Dietrich S 2017 {\em Active
colloids in the context of chemical kinetics} J. Phys. A: Math. Theor. {\bf
50} 134001

\bibitem{dist5} Grebenkov DS, Metzler R and Oshanin G 2017 {\em Effects
of the target aspect ratio and intrinsic reactivity onto diffusive search
in bounded domains} New J. Phys. {\bf 19} 103025

\bibitem{dist7} Grebenkov DS, Metzler R, Oshanin G, Dagdug L, Berezhkovskii
AM and Skvortsov AT 2019 {\em Trapping of diffusing particles by periodic
absorbing rings on a cylindrical tube} J. Chem. Phys. {\bf 150} 206101

\bibitem{dist2} Grebenkov D, Metzler R and Oshanin G 2019 {\em Full
distribution of first exit times in the narrow escape problem}  New
J. Phys. {\bf 21} 122001

\bibitem{dist4} Grebenkov D, Metzler R and Oshanin 2018 {\em Towards a full
quantitative description of single-molecule reaction kinetics in biological
cells}  Phys. Chem. Chem. Phys. {\bf 20} 16393

\bibitem{ol1} B\'enichou O, Moreau M and Oshanin G 2000 {\em Kinetics of
stochastically gated diffusion-limited reactions and geometry of random walk
trajectories} Phys. Rev. E {\bf 61} 3388

\bibitem{Collins49}   Collins FC and Kimball GE 1949 {\em Diffusion-controlled
reaction rates} J. Coll. Sci. {\bf 4} 425

\bibitem{lawley} Lawley SD and Miles CE 2019 {\em Diffusive search for
diffusing targets with fluctuating diffusivity and gating} J. Nonlinear
Sci. {\bf 29} 2955

\bibitem{lawley2}  Lawley SD and Keener JP 2015 {\em A New Derivation
of Robin Boundary Conditions through Homogenization of a Stochastically
Switching Boundary} SIAM J. Appl. Dyn. Sys.  {\bf 14} 1845-1867

\bibitem{denis} Grebenkov DS {\em Imperfect diffusion-controlled reactions},
in Ref. \cite{redner2}, Ch. $8$

\bibitem{Grebenkov20}  Grebenkov DS 2020 {\em Paradigm Shift in
Diffusion-Mediated Surface Phenomena} Phys. Rev. Lett. {\bf 125} 078102

\bibitem{smoluchowski} v Smoluchowski M 1916 {\em Drei Vortr\"age \"uber
Diffusion, Brownsche Molekularbewegung und Koagulation von Kolloidteilchen},
Z. Phys. {\bf 17}, 557

\bibitem{race} Grebenkov DS, Metzler R and Oshanin G 2021 {\em A molecular
relay race: sequential first-passage events to the terminal reaction centre
in a cascade of diffusion controlled processes} New J. Phys. {\bf 23}  093004


{\clr
\bibitem{Sapoval94}  Sapoval B (1994)
{\em General Formulation of Laplacian Transfer Across Irregular Surfaces}
Phys. Rev. Lett. {\bf 73} 3314 

\bibitem{Sapoval02}  Sapoval B, Filoche M, and  Weibel E (2002)
{\em Smaller is better -- but not too small: A physical scale for the design of the mammalian pulmonary acinus}
Proc. Nat. Acad. Sci. USA {\bf 99} 10411

\bibitem{Grebenkov03}  Grebenkov DS, Filoche M, and Sapoval B (2003)
{\em Spectral Properties of the Brownian Self-Transport Operator} Eur. Phys. J. B {\bf 36} 221-231 

\bibitem{Sapoval05}	Sapoval B, Andrade JS Jr, Baldassari A, Desolneux A, Devreux F, Filoche M, Grebenkov DS, Russ S (2005)
{\em New Simple Properties of a Few Irregular Systems} Physica A {\bf 357} 1-17

\bibitem{Grebenkov06c}   Grebenkov DS (2006) 
{\em Scaling Properties of the Spread Harmonic Measures} Fractals {\bf 14} 231-243

\bibitem{Grebenkov15c}	Grebenkov DS (2015)
{\em Analytical representations of the spread harmonic measure} Phys. Rev. E {\bf 91} 052108
}



\bibitem{Arendt14} Arendt W A. ter Elst AFM, Kennedy JB and Sauter M 2014
{\em The Dirichlet-to-Neumann operator via hidden compactness}
J. Funct. Anal. {\bf 266} 1757-1786

\bibitem{Daners14} Daners D 2014 {\em Non-positivity of the semigroup
generated by the Dirichlet-to-Neumann operator} Positivity {\bf 18} 235-256

\bibitem{Arendt15} Arendt W and ter Elst AFM 2015 {\em The
Dirichlet-to-Neumann Operator on Exterior Domains} Potential Anal. {\bf
43} 313-340

\bibitem{Hassell17} Hassell A and Ivrii V 2017{ \em Spectral asymptotics
for the semiclassical Dirichlet to Neumann operator}, J. Spectr. Theory {\bf
7} 881-905

\bibitem{Girouard17} Girouard A and Polterovich I 2017 {\em Spectral
geometry of the Steklov problem} J. Spectr. Theory {\bf 7} 321-359

\bibitem{Grebenkov20c}  Grebenkov DS 2020 {\em Surface Hopping Propagator:
An Alternative Approach to Diffusion-Influenced Reactions}, Phys. Rev. E
{\bf 102} 032125

\bibitem{Grebenkov20d}  Grebenkov DS 2020 {\em Joint distribution of multiple
boundary local times and related first-passage time problems with multiple
targets} J. Stat. Mech. 103205

\bibitem{dist1} Grebenkov D, Metzler R and Oshanin G 2018 {\em Strong
defocusing of molecular reaction times results from an interplay of geometry
and reaction control} Comm. Chem. {\bf 1}  96

\bibitem{aljaz} Godec A and Metzler R 2015 \emph{Optimization and universality
of Brownian search in quenched heterogeneous media} Phys. Rev. E \textbf{91}
052134

\bibitem{aljazprx} Godec A and Metzler R 2016 \emph{Universal proximity effect
in target search kinetics in the few encounter limit} Phys. Rev. X \textbf{6}
041037

\bibitem{aljaz1} Godec A and Metzler R 2017 \emph{First passage time statistics for
two-channel diffusion}, J. Phys. A \textbf{50} 084001

{\clr
\bibitem{Benichou10f}   B\'enichou O, Chevalier C, Klafter J, Meyer B, and  Voituriez R (2010)
{\em Geometry-controlled kinetics}, Nature Chem. {\bf 2} 472-477 
}

\bibitem{Reva21} Reva M, DiGregorio DA, and Grebenkov DS 2021 \emph{A
first-passage approach to diffusion-influenced reversible binding: insights
into nanoscale signaling at the presynapse}, Sci. Rep. \textbf{11} 5377 


\bibitem{24} Singer A, Schuss Z, Holcman D and Eisenberg B 2006 {\it Narrow
Escape I} J. Stat. Phys.  {\bf 122} 437

\bibitem{25} Singer A, Schuss Z and Holcman D 2006 {\it Narrow Escape II}
J. Stat. Phys.  {\bf 122} 465

\bibitem{26} Singer A, Schuss Z and Holcman D 2006 {\it Narrow Escape III}
J. Stat. Phys.  {\bf 122} 491

\bibitem{Rupprecht17} Grebenkov DS and Rupprecht J-F 2017 \emph{The escape
problem for mortal walkers}, J. Chem. Phys. {\bf 146} 084106

\bibitem{yu} Yu J, Xiao J, Ren X, Lao K and Xie XS 2006 {\em Probing Gene
Expression in Live Cells, One Protein Molecule at a Time} Science \textbf{311}
1600

\bibitem{elf} Elf J, Li G-W and Xie XS 2007 {\em Probing Transcription Factor
Dynamics at the Single-Molecule Level in a Living Cell}, Science \textbf{316},
1192

\bibitem{wang} Wang M, Zhang J, Xu H and Golding I 2019 {\em Measuring
transcription at a single gene copy reveals hidden drivers of bacterial
individuality}, Nature Microbiol. \textbf{4} 2118

\bibitem{weigel} Weigel AV, Tamkun MM and Krapf D 2013 {\em Quantifying the
dynamic interactions between a clathrin-coated pit and cargo molecules},
Proc. Natl. Acad. Sci. USA \textbf{110} E4591

\bibitem{kepes} K{\'e}p{\`e}s F 2004 {\em Periodic Transcriptional Organization of
the E. coli Genome}, J. Mol. Biol. \textbf{340}

\bibitem{kolesov} Kolesov G, Wunderlich Z, Laikova ON, Gelfand MS and
Mirny LA 2007 {\em How gene order is influenced by the biophysics
of transcription regulation}, Proc. Natl. Acad. Sci. USA \textbf{104} 13948

\bibitem{otto} Pulkkinen O and Metzler R 2013 {\em Distance matters: the impact of
gene proximity in bacterial gene regulation}, Phys. Rev. Lett. \textbf{110} 198101

\bibitem{otto1} Pulkkinen O and Metzler R 2015 \emph{Variance-corrected
Michaelis-Menten equation predicts transient rates of single-enzyme reactions
and response times in bacterial gene regulation}, Sci. Rep. \textbf{5} 17820

\bibitem{kuehn} K{\"u}hn T, Ihalainen TO, Hyv{\"a}luoma J, Dross N, Willman SF,
Langowski J, Vihinen-Rante M and Timonen J 2011 {\em Protein Diffusion in Mammalian
Cell Cytoplasm}, PLoS ONE \textbf{6} e22962

\bibitem{han} Han D, Korabel N, Chen R, Johnston M, Gavrilova A, Allan VJ,
Fedotov S and Wigh TA 2020 {\em Deciphering anomalous heterogeneous
intracellular transport with neural networks}, eLife \textbf{9}, e52224

\bibitem{ma} Ma J, Do M, Le Gros MA, Peskin CS, Larabell CA, Mori Y and
Isascson SA 2020 {\em Strong intracellular signal inactivation produces sharper
and more robust signaling from cell membrane to nucleus}, PLoS Comp. Biol.
\textbf{16} e1008356

\bibitem{Doris}  Witzel P, G\"otz M, Lanoisel\'ee Y, Franosch T, Grebenkov DS, and Heinrich D (2019)
{\em Heterogeneities Shape Passive Intracellular Transport} Biophys. J. {\bf 117} 203-213


\bibitem{N5}  Grebenkov D, Metzler R and Oshanin G  2020 {\em From
single-particle stochastic kinetics to macroscopic reaction rates: fastest
first-passage time of $N$ random walkers} New J. Phys. {\bf 22} 103004



\bibitem{bray} Bray AJ and Blythe RA 2002  {\em Exact Asymptotics for
One-Dimensional Diffusion with Mobile Traps} Phys. Rev. Lett. {\bf 89} 150601

\bibitem{osh} Oshanin G, B\'enichou O, Coppey M and Moreau M  2002 {\em
Trapping reactions with randomly moving traps: Exact asymptotic results for
compact exploration} Phys. Rev. E {\bf 66} 060101(R)

\bibitem{pascal} Moreau M, Oshanin G, B\'enichou O and Coppey M 2003 {\em
Pascal principle for diffusion-controlled trapping reactions} Phys. Rev. E
{\bf 67} 045104(R); 2004 {\em Lattice theory of trapping reactions with
mobile species}  Phys. Rev. E {\bf 69} 046101

\bibitem{satya} Bray AJ, Majumdar SN and Blythe RA 2003 {\em Formal solution of
a class of reaction-diffusion models: Reduction to a single-particle problem}
Phys. Rev. E {\bf 67} 060102(R)

\bibitem{osha} Yuste SB, Oshanin G, Lindenberg K, B\'enichou O and Klafter
J  2008 {\em  Survival probability of a particle in a sea of mobile traps:
A tale of tails} Phys. Rev. E {\bf 78}  021105

\bibitem{LeVot}  Le Vot F, Yuste SB, Abad E and Grebenkov DS 2020 {\em
First-encounter time of two diffusing particles in confinement} Phys. Rev. E
{\bf 102} 032118

\bibitem{N1} Meerson B and Redner S 2015 {\em Mortality, redundancy, and
diversity in stochastic search} Phys. Rev. Lett. {\bf 114} 198101

\bibitem{N2} Reynaud K, Schuss Z, Rouach N and Holcman D 2015 {\em Why so
many sperm cells?} Commun. Integr. Biol. {\bf 8} e1017156

\bibitem{N3} Lawley S D and Madrid J B 2020 {\it A probabilistic approach to
extreme statistics of Brownian escape times in dimensions $1$, $2$, and $3$}
J. Nonlinear Sci. {\bf 30} 1207

\bibitem{N4} Lawley S D 2020 {\it Distribution of extreme first passage
times of diffusion} J. Math. Biol. {\bf 80} 2301

\bibitem{Grebenkov19d} Grebenkov DS 2019 \emph{A unifying approach to first-passage 
time distributions in diffusing diffusivity and switching diffusion models}, J.
Phys. A \textbf{52} 174001 

\bibitem{Lanoiselee18} Lanoisel\'ee Y, Moutal N, and Grebenkov DS (2018) 
\emph{Diffusion-limited reactions in dynamic heterogeneous media}, Nature
Commun. {\bf 9} 4398


\bibitem{Grebenkov19c}  Grebenkov DS 2019 {\em Probability distribution of
the boundary local time of reflected Brownian motion in Euclidean domains}
Phys. Rev. E {\bf 100} 062110



\end{thebibliography}
\end{document}